\documentclass[%
twocolumn,
superscriptaddress,
amsmath,
amssymb,
aps,
prb,
floatfix,
]{revtex4-2}
\usepackage{graphicx}%
\graphicspath{{figures/}}
\usepackage{tikz}
\usetikzlibrary{arrows.meta}
\usetikzlibrary{calc}
\usepackage{amsmath,amssymb,amsfonts,physics}%
\usepackage{amsthm}%
\usepackage{xcolor}%
\usepackage{hyperref}
\usepackage[capitalise]{cleveref}
\usepackage{glossaries}
\newacronym{DMRG}{DMRG}{density-matrix renormalization group}
\newacronym{AFM}{AFM}{antiferromagnet}
\newacronym{BCS}{BCS}{Bardeen-Cooper-Schrieffer}
\newacronym[shortplural={MPS}]{MPS}{MPS}{matrix-product states}
\newacronym{CTKS}{CTKS}{complex-time Krylov space}
\newacronym{ARPES}{ARPES}{angle-resolved photoemission spectroscopy}

\begin{document}

\title[Article Title]{Two-channel physics in a lightly doped antiferromagnetic Mott insulator\\ revealed by two-hole spectroscopy}

\author{Pit Bermes}
\email{pit.bermes@lmu.de}
\affiliation{%
 Department of Physics and Arnold Sommerfeld Center for Theoretical Physics (ASC),
Ludwig-Maximilians-Universit\"at M\"unchen, Theresienstr. 37, M\"unchen D-80333, Germany
}%
\affiliation{%
 Munich Center for Quantum Science and Technology (MCQST), Schellingstr. 4, D-80799 M\"unchen, Germany
}%

\author{Sebastian Paeckel}
\affiliation{%
 Department of Physics and Arnold Sommerfeld Center for Theoretical Physics (ASC),
Ludwig-Maximilians-Universit\"at M\"unchen, Theresienstr. 37, M\"unchen D-80333, Germany
}%
\affiliation{%
 Munich Center for Quantum Science and Technology (MCQST), Schellingstr. 4, D-80799 M\"unchen, Germany
}%

\author{Annabelle Bohrdt}
\affiliation{%
Department of Physics and Arnold Sommerfeld Center for Theoretical Physics (ASC),
Ludwig-Maximilians-Universit\"at M\"unchen, Theresienstr. 37, M\"unchen D-80333, Germany
}%
\affiliation{%
 Munich Center for Quantum Science and Technology (MCQST), Schellingstr. 4, D-80799 M\"unchen, Germany
}%

\author{Lukas Homeier}
\affiliation{%
 JILA and Department of Physics, University of Colorado, Boulder, CO, 80309, USA
}%
\affiliation{%
 Center for Theory of Quantum Matter, University of Colorado, Boulder, CO, 80309, USA
}%

\author{Fabian Grusdt}%
\email{fabian.grusdt@lmu.de}
\affiliation{%
 Department of Physics and Arnold Sommerfeld Center for Theoretical Physics (ASC),
Ludwig-Maximilians-Universit\"at M\"unchen, Theresienstr. 37, M\"unchen D-80333, Germany
}%
\affiliation{%
 Munich Center for Quantum Science and Technology (MCQST), Schellingstr. 4, D-80799 M\"unchen, Germany
}%

\date{\today}

% \showthe\columnwidth

\begin{abstract}
Understanding pairing in the strong-coupling regime of doped Mott insulators remains an open problem in the context of cuprate superconductors. We perform ultra-high resolution numerical simulations of spectral functions in the highly underdoped $t-J$ model and discover two coupled branches of hole pairs emerging at low energies in the largely unexplored two-particle spectrum. As spin anisotropy is tuned from the Ising limit to the $SU(2)$-symmetric Heisenberg regime, the lowest $d$-wave pair evolves from a single bipolaronic branch into two hybridized branches separated by an avoided crossing. We explain this behaviour using an effective two-channel model involving a tightly bound bipolaronic state and a second channel associated with two magnetic polarons. The model reproduces the qualitative low-energy spectra and implies near-resonant $d$-wave interactions in the $SU(2)$-symmetric $t-J$ model, consistent with proximity to an emergent Feshbach-type resonance. To probe these predictions experimentally, we propose a Raman spectroscopy scheme for the attractive Hubbard model that can be directly implemented using ultracold atoms in optical lattices. Our work establishes two-particle spectroscopy, beyond single-particle Green's functions, as a powerful tool for revealing the microscopic origins of unconventional superconductivity. 
\end{abstract}

\keywords{Hubbard model, t-J model, doped Mott insulator, magnetic polarons}

\maketitle

% % % % % % % % % % % % % % % % % % % % % % % % % % % % % % 
%\section{Introduction}
%\label{sec:introduction}
% % % % % % % % % % % % % % % 
Strong electronic correlations, as broadly captured by $t-J$ and Hubbard models~\cite{Lee2006}, are widely believed to underlie high-$T_c$ superconductivity discovered in cuprate compounds~\cite{Bednorz1986}. Nevertheless, the microscopic nature of the superconducting state as well as its relation to the pseudogap phase and the associated pairing fluctuations~\cite{Uemura1991,Emery1995,niu2024arXiv}, remain poorly understood. On the one hand, it is an experimentally established fact that its pairing symmetry is of $d_{x^2-y^2}$ form~\cite{Wollman1993,Hashimoto2014}; and such $d$-wave fluctuations constitute the dominant pairing fluctuations found in numerical simulations of $t-J$ and Hubbard models~\cite{Dagotto1990pair,Corboz2014,Xu2024}. Moreover, at high doping or in the limit of weak Hubbard interactions, the $d$-wave channel is theoretically established to be the dominant instability of the Fermi-surface~\cite{Scalapino1987,Halboth2000}. On the other hand, understanding the fate of the superconductor at large Hubbard-$U$ interactions and low doping -- referred to as the strong coupling regime in the following -- continues to represent an open problem. This is also where the highest superconducting $T_c$ are realized. 

A popular point of view stipulates that the $d$-wave superconductor realized at strong coupling is an adiabatic extension of the weakly coupled \gls{BCS} state. Extensive studies based on the functional renormalization group~\cite{Metzner2012, Vilardi2019}  confirm this view. Nevertheless, it has also been established that the superconducting state at strong coupling has some characteristics well beyond what can be explained by a mean-field \gls{BCS} state~\cite{Stajic2003}: it is destroyed by phase rather than pairing fluctuations~\cite{Uemura1991,Emery1995,Corson1999,Zhou2019,niu2024arXiv}; its specific heat deviates significantly from \gls{BCS} expectations~\cite{Loram1994}; and its coherence length (related to the size of a Cooper pair) is small, on the order of a lattice constant, and increases away from optimal doping~\cite{Sonier2007,Wen2003}. Here we show that pairing in the strong-coupling regime exhibits a characteristic two-channel structure consistent with proximity to a Feshbach-type resonance, shedding new light on the observed non-\gls{BCS} characteristics.

Remarkably little is known directly about the microscopic structure of the involved pairs in high-$T_c$ superconductors, beyond their $d$-wave nature. In contrast to the one-electron Green's function, which has been excessively studied theoretically~\cite{Sordi2012,Schaefer2021,Simkovic2024} and can be directly accessed in high-resolution \gls{ARPES}~\cite{Damascelli2003,Shen2005,Chen2019,Kunisada2020}, the two-particle Green's function, containing information about the structure of pairs, remains largely unexplored. This is partly due to the challenges associated with accurate numerical computations, as well as direct measurements in solids. The primary tool for accessing properties of pairs so far are transport experiments~\cite{Scalapino1970,Anderson1972,Bergeal2008}, including in particular shot noise measurements~\cite{Zhou2019,niu2024arXiv,Bastiaans2021}, which cannot fully reveal the structure and characteristics of the pairs.

\begin{figure}[t]
    \includegraphics[width=0.99\linewidth]{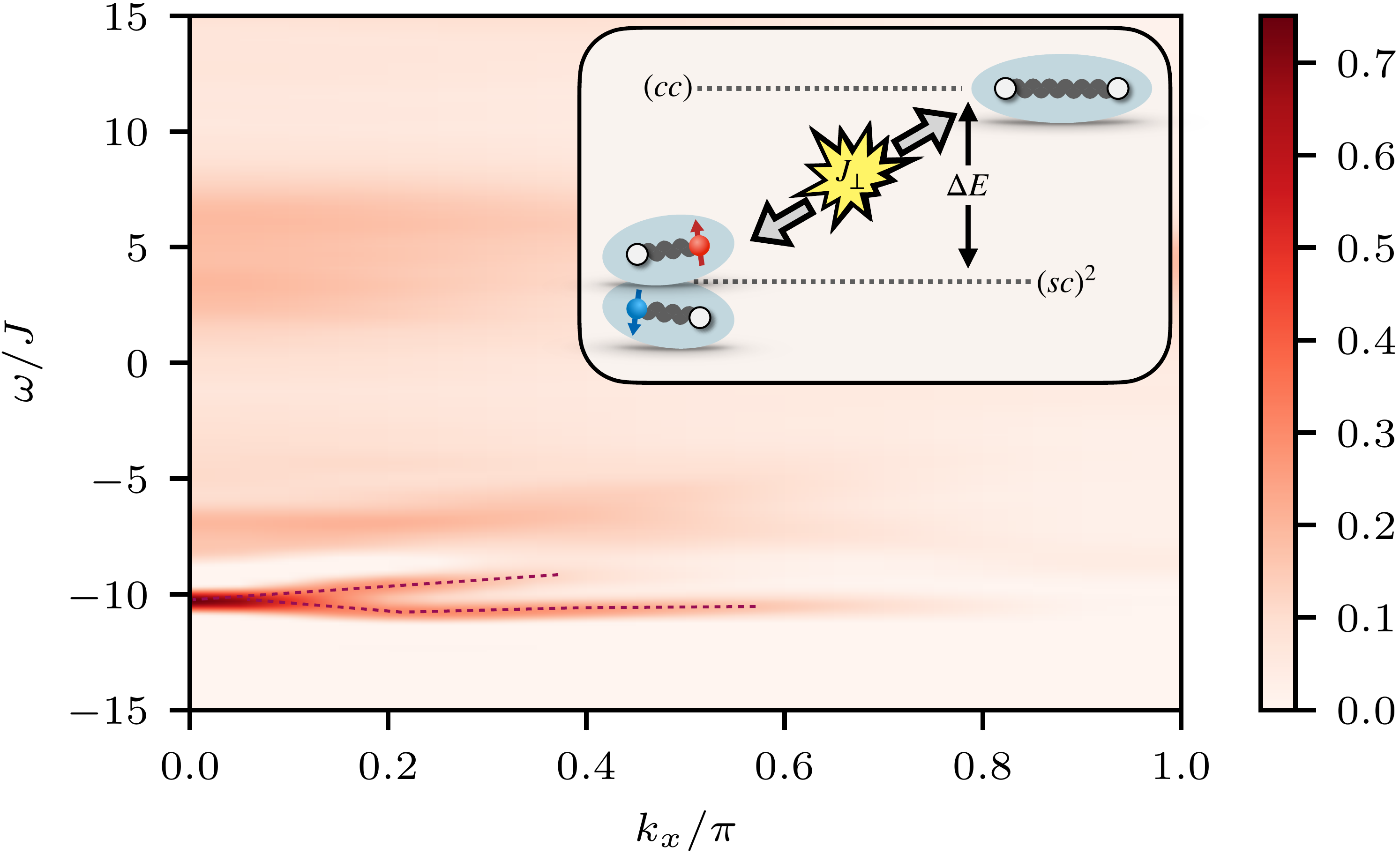}
    \caption{Signatures of two-channel physics in the two-hole rotational spectrum with $d$-wave symmetry, $A^{(2)}(\mathbf{k},\omega)$, starting from an initially undoped Heisenberg \gls{AFM}, computed at fixed $k_y=\pi/2$ using \acrshort{MPS}~\cite{Bohrdt2023}. Dashed lines are guides to the eye tracking local maxima in the low-energy part of the spectrum to highlight the two hybridized branches we find. The inset illustrates the effective two-channel model~\cite{Homeier2025} and its ingredients -- magnetic polaron (sc) and bipolaron (cc) channels -- which we use to explain the observed level splitting. These numerical simulations were performed in Ref.~\cite{Bohrdt2023} on a $40 \times 4$-cylinder, at $t/J=3$, using time-dependent \glspl{MPS}~\cite{Zaletel2015}.}
    \label{fig:overview}
\end{figure}

Here we perform a numerical study of two-hole spectra in the $t-J$ model. The ultra-high frequency resolution we achieve leads us to discover a previously unknown avoided level crossing between two distinct paired states of two holes doped into a Heisenberg \gls{AFM}, see \cref{fig:overview}. This indicates near-resonant, attractive interactions in the strong-coupling regime of the $t-J$ model, characteristic for a system in the vicinity of a Feshbach resonance~\cite{Feshbach1958,Chen2024RMP,Homeier2025}. Specifically, our results show how the bipolaronic resonance of two holes in a N\'eel state, identified in Refs.~\cite{Bohrdt2023,Grusdt2023}, hybridizes with a state consisting of two weakly bound individual holes, or magnetic polarons. We propose a direct measurement scheme for cold atom experiments and explain our numerical findings by a two-channel model~\cite{Homeier2024, Homeier2025} of mesonic charge carriers~\cite{Beran1996,Grusdt2018,Bohrdt2021,Grusdt2023,Bohrdt2023} derived in the strong-coupling regime. The latter suggests that hole-doped cuprates are on the \gls{BCS}-side of -- but  close to -- an emergent Feshbach resonance~\cite{Homeier2025}, in line with recent theoretical and experimental evidence~\cite{Sous2023,Chen2024test}. 

To explore the influence of the underlying \gls{AFM} order, we consider a model Hamiltonian with a tunable easy-axis anisotropy, $J_\perp \leq J_z$,
\begin{align}
    \hat{\mathcal{H}}
    &=
    -t \sum_{\langle \mathbf i,\mathbf j\rangle,\sigma}
    \hat{\mathcal{P}}
    \left(
        \hat c^\dagger_{\mathbf i,\sigma}\hat c^{\vphantom{\dagger}}_{\mathbf j,\sigma} + \mathrm{h.c.}
    \right) 
    \hat{\mathcal{P}}
    -
    \frac{J_z}{4}\sum_{\langle \mathbf i,\mathbf j\rangle} \hat n_\mathbf{i}\hat n_\mathbf{j}
    \notag \\
    &\phantom{=-}
    +\sum_{\langle \mathbf i,\mathbf j\rangle}
    \left(
        J_\perp (\hat S^x_\mathbf{i}\hat S^x_\mathbf{j} + \hat S^y_\mathbf{i}\hat S^y_\mathbf{j})
        +
        J_z \hat S^z_\mathbf{i}\hat S^z_\mathbf{j}
    \right) \;,
    \label{eq:hamiltonian}
\end{align}
where $\hat{\mathcal{P}}$ projects to maximally singly-occupied sites; $\hat{c}^\dagger_{\mathbf i,\sigma}$ creates a fermion of spin $\sigma$ at site $\mathbf{i}$, and $\hat{\mathbf{S}}_{\mathbf{j}}$ ($\hat{n}_{\mathbf{j}}$) denotes the corresponding spin (density). Our study of two-hole spectra in this model, on numerically accessible four-leg cylinders, leverages ultra-high frequency resolution at low energies, achieved through a numerical \gls{MPS}-based technique~\cite{Paeckel2024} beyond established schemes~\cite{PAECKEL2019}, see Methods Sec.~\ref{secMethods}.

\begin{figure*}[ht]
    \centering
    \includegraphics[width=0.99\linewidth]{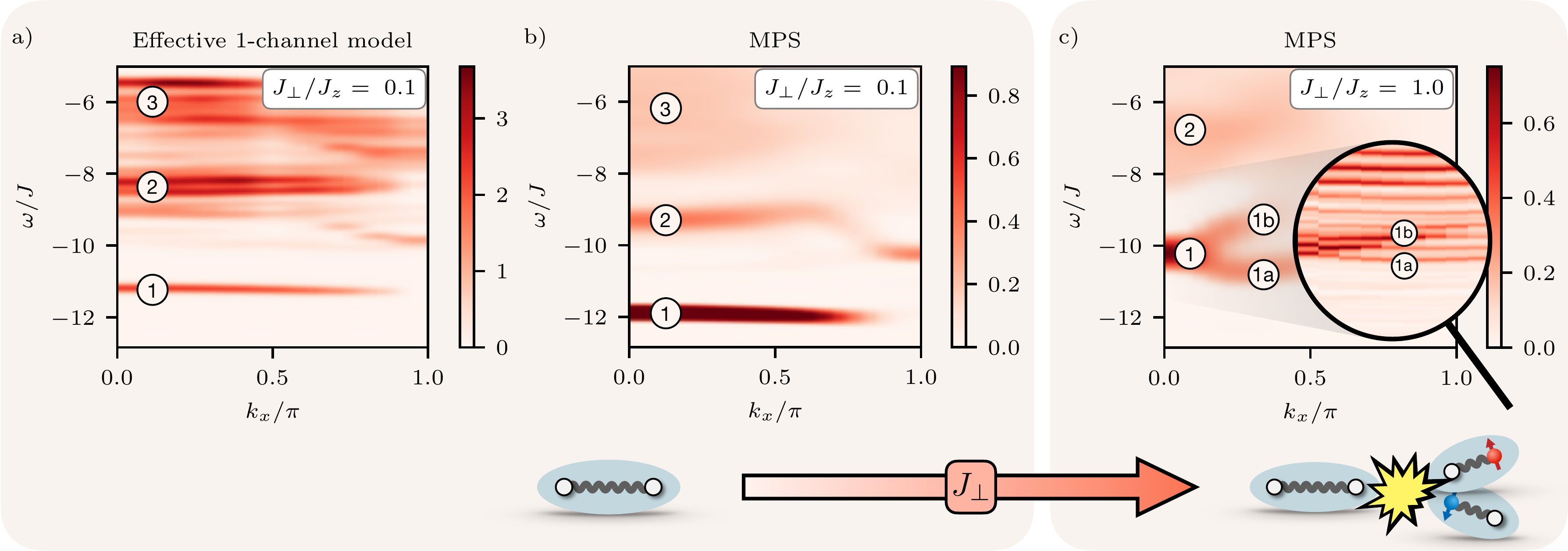}
    \caption{Emergence of two paired branches in the two-hole spectrum $A^{(2)}(\mathbf{k}, \omega)$ of the $t-J$ model. In a), b) we consider an easy-axis \gls{AFM}, $J_\perp = 0.1 J$ ($J_z=J$). The full numerical \gls{MPS} spectrum~\cite{Bohrdt2023} shown in b) is in excellent qualitative agreement with the effective single-channel model a), based on two holes connected by a string of displaced spins. In particular, all observed branches -- labeled \textcircled{1} - \textcircled{3} -- are found to be in one-to-one correspondence. The effective model in a) describes only the tightly-bound bipolaronic pairing channel, without coupling to the magnetic polarons; it has been artificially broadened for better comparison. In c) we increase $J_\perp = J_z =J$ and show \gls{MPS} spectra in the $SU(2)$ invariant $t-J$ model~\cite{Bohrdt2023}, which we compare to ultra-high resolution spectra obtained by the \glslink{CTKS}{\glsentryshort{CTKS}} approach described in the methods (inset). The lowest pair resonance, \textcircled{1}, is found to split into two branches, \textcircled{1a} and \textcircled{1b} -- indicating the presence of an additional paired eigenstate hybridizing with the bipolaronic hole pair. We work at fixed $k_y=\pi/2$ and in the d-wave channel, $m_4 = 2$, at $t/J=3$.
    }
    \label{fig:jperp-scan}
\end{figure*}

%%%%%%
\section{Numerical signatures of two coupled pairing channels}\label{sec:numerical-signatures}
%%%%%%
\emph{Two-hole spectrum.--}
In order to probe the internal structure of hole pairs, we start from an undoped \gls{AFM} ground state $\ket{\psi_0}$ and compute the two-hole spectrum $A^{(m_4)}(\mathbf{k},\omega) = - \pi^{-1} {\rm Im} \mathcal{G}^{(m_4)}(\mathbf{k},\omega)$. The latter is related to the retarded two-particle Green's function~\cite{Bohrdt2023}
\begin{equation}
 \mathcal{G}^{(m_4)}(\mathbf{j},t) = \langle\Psi_0|\hat{\Delta}^{(\textrm{s})\dagger}(\mathbf{j}, t; m_4)
    \hat{\Delta}^{(\textrm{s})}(\mathbf{0}, 0; m_4)|\Psi_0\rangle\,,
 \label{eqDefG}
\end{equation}
describing pairs of holes with $m_4 \in \{0,1,2,3\}$ units of $C_4$ relative angular momentum; here 
\begin{equation}
\label{eq:pair-creation-operator}
    \hat{\Delta}^{(\textrm{s})}(\mathbf{j}; m_4) = \sum_{\mathbf{i}:\langle\mathbf{i},\mathbf{j}\rangle}e^{im_4\varphi_{\mathbf{i} - \mathbf{j}}}(\hat{c}_{\mathbf{i},\downarrow}\hat{c}_{\mathbf{j},\uparrow} - \hat{c}_{\mathbf{i},\uparrow}\hat{c}_{\mathbf{j},\downarrow})\,,
\end{equation}
creates a spin-singlet hole pair on nearest-neighbor sites, with $\varphi_{\mathbf{i} - \mathbf{j}}$ the polar angle of $\mathbf{i} - \mathbf{j}$. 

The two-hole spectrum $A^{(m_4)}(\mathbf{k},\omega)$ that we compute characterizes eigenstates of the model Eq.~\eqref{eq:hamiltonian} with two doped holes, at energy $\omega$ and momentum $\mathbf{k}$ defined relative to the undoped ground state $\ket{\psi_0}$. The spectral weight describes the size of the overlap of such eigenstates with the local pair $ |\varphi^{m_4}_\mathbf{j} \rangle = \hat{\Delta}^{(\textrm{s})}(\mathbf{j}; m_4) \ket{\psi_0}$. In addition to a continuum of unbound two-hole states at high energies, the pair spectrum computed with \gls{MPS} contains a set of narrow resonances (bands) with a well-defined energy-momentum relation at low energies~\cite{Bohrdt2023}, see \cref{fig:overview}. These correspond to discrete bound states of the two holes, and partly exist in a range of energies above the threshold where decay into two independent individual holes, or magnetic polarons, is possible. This is expected to lead to a non-vanishing spectral width (i.e., a finite lifetime) of the resonances, below or comparable to our numerical frequency resolution in \cref{fig:overview}. 

We go beyond previous studies of the two-hole spectrum~\cite{Bohrdt2023} by analyzing its low-energy features in greater detail. To this end we performed new \gls{MPS} simulations with a more than an order of magnitude improvement in energy resolution, achieved by \gls{CTKS} expansion~\cite{Paeckel2024} and explained in the Methods Sec.~\ref{secMethods}. A first example is shown in the inset of \cref{fig:jperp-scan} c), which resolves the low-energy features in the same parameter regime as \cref{fig:overview}. Here the narrowly spaced, discrete lines correspond to finite-size splittings associated with a narrow four-leg cylinder. Overall, our new higher-resolution numerics and systematic parameter exploration establish the low-energy double-peak feature, previously hinted at in the lower-resolution \gls{MPS} results of \cref{fig:overview}, as a genuine and physically robust characteristic of the spectrum.

\emph{Beyond one-channel physics.--}
Now we commence our analysis of the physics associated with the two emerging pair branches revealed in our numerical simulations of the $t-J$ model in \cref{fig:overview}. To this end we start by discussing the two-hole spectrum in a strongly anisotropic XXZ \gls{AFM}, where $J_\perp \ll J_z$, whose undoped ground state is well described by an easy-axis N\'eel state. In this case, all long-lived pair resonances can be explained by an effective single-channel meson picture, where the two holes are tightly bound into a bipolaron by a string of displaced spins in the N\'eel background~\cite{Grusdt2023}. This is shown by an explicit comparison of string and \gls{MPS} spectra at $J_\perp = 0.1 J_z$ in \cref{fig:jperp-scan} a)-b), sharing the same qualitative band structure in the low-energy regime (several units of $J$ wide), including e.g. a pronounced redistribution of spectral weight near $k_x = \pi$.

In the single-channel meson picture we employ, a mobile hole leaves behind a string of displaced strings which is retraced by the second hole, thereby re-establishing an undisturbed N\'eel background~\cite{Shraiman1988a,Grusdt2023}. A linear confining force leads to a tightly-bound bipolaronic state, whose internal ro-vibrational excitations give rise to the higher bands observed in \cref{fig:jperp-scan} a)-b). Our corresponding calculations in \cref{fig:jperp-scan} a) are based on a truncated-basis Krylov space approach~\cite{Vidmar2013}, described in detail in Ref.~\cite{Homeier2024}. They include loop effects and self-interactions of strings, thus going beyond the simplified linear-string theory calculations performed in Ref.~\cite{Grusdt2023}. 

We proceed by analyzing the spectrum of the $SU(2)$ invariant $t$-$J$ model, $J_\perp = J_z = J$. In \cref{fig:jperp-scan} c) its low-energy part is shown, computed by \gls{MPS}~\cite{Bohrdt2023}, for the same momentum cut and rotational sector as in the XXZ \gls{AFM} shown in b). Up to an overall shift to higher energies and increased line broadening of the higher lying states, the bipolaronic bands previously matched with the single-channel meson spectrum can still be identified (features \textcircled{1} and \textcircled{2} in the figure). However, a striking deviation from the single-channel structure is observed in the lowest band \textcircled{1}, which splits into two weakly dispersive bands \textcircled{1a} and \textcircled{1b}. Based on our one-to-one matching of the other features in the two-hole spectrum with the effective single-channel meson model, we conclude that the observed splitting of the lowest band into two branches must be due to the emergence of a second scattering channel in the Heisenberg limit $J_\perp \to J_z$. 

\begin{figure}[t]
    \centering
    \includegraphics{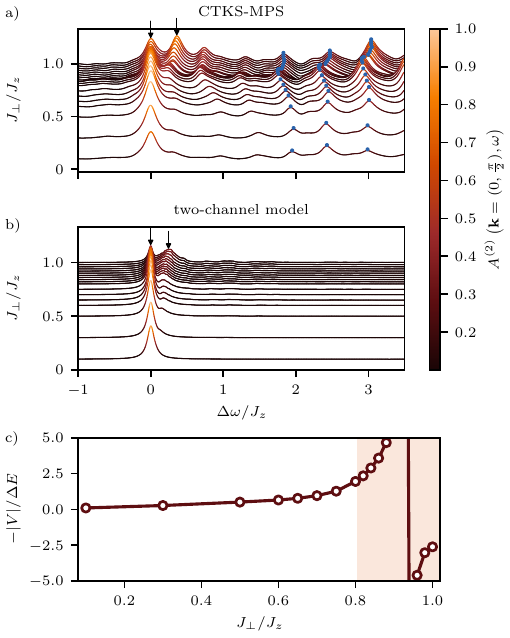}
    \caption{Emergent scattering resonance between low-energy two-hole states. a) We perform high-resolution \gls{MPS} simulations of the two-hole spectrum $A^{(2)}(\mathbf{k}, \omega)$ on a $40 \times 4$ cylinder using the \gls{CTKS} method. We fix $\mathbf{k}=(0,\pi/2)$, $t/J=3$ and tune $J_\perp / J_z$, indicated by baseline offsets of the curves along the $y$-axis. Around $J_\perp/J_z \approx 0.8$, the shoulder of the lowest-energy peak develops into an avoided level crossing of two paired states, highlighted by small arrows. Energies $\Delta \omega$ are plotted with respect to the lowest quasi-particle peak, and spectra are normalized to their maximal value. The blue dots mark some of the peaks to highlight the kink of the higher peak positions.
    b) The avoided level crossing is captured by an effective two-channel model, showing only the lowest-lying branches. The single fit parameter, $\Delta E$, in the two-channel model is obtained by comparison to the \gls{MPS} simulations in a). These results can be interpreted as an emergent Feshbach resonance between two branches of paired states: In panel c), we show the ratio of the coupling, $V$, between the two channels, and $\Delta E$, the fitted bare energy offset of the effective two-channel model. $-|V|^2/\Delta E \propto g_d$ is proportional to the effective Feshbach interaction between two magnetic polarons, mediated by the coupling to the tightly-bound bipolaronic pair. The shaded region marks the strong coupling regime, i.e., the location of the emergent Feshbach resonance, where the hybridization $V>2\Delta E$ exceeds the offset.}
    \label{fig:jp-scan}
\end{figure}

To validate this picture, we continuously tune $J_\perp/J_z$ from the single-channel Ising limit to the Heisenberg regime, and track the evolution of the two-hole spectrum. Our results, leveraging the ultra-high frequency resolution of the~\gls{CTKS} method, are shown in \cref{fig:jp-scan} a), focusing on the low-energy regime. To highlight the tell-tale signs of an avoided level crossing, frequencies $\Delta \omega$ are plotted relative to the lowest-lying resonance. 

As the transverse spin coupling $J_\perp$ is increased we observe how a small shoulder above the lowest peak emerges. Around $J_\perp/J_z \approx 0.8$, this shoulder develops into a separate peak to which significant spectral weight is transferred, mostly by a reduction of weight in the lowest-lying resonance. In contrast, the structure of the higher-lying peaks remains largely unchanged (see \cref{supp:CTKS} for a plot of a larger frequency window). 

%At the same value $J_\perp/J_z \approx 0.8$, we observe a pronounced kink of all higher peak positions $\Delta \omega_n = \omega_n - \omega_0$ in \cref{fig:jp-scan} a), with $\partial_{J_\perp}\Delta \omega_n$ changing sign around $J_\perp/J_z \approx 0.8$.
In addition, we track the positions of the higher peaks $\Delta \omega_n = \omega_n - \omega_0$ (blue dots in \cref{fig:jp-scan}). As we increase $J_\perp$, these peaks initially shift to lower energies, exhibiting a sharp kink at around the same value $J_\perp/J_z \approx 0.8$, beyond which their energy increases again.
We attribute this behavior to a kink in just the first peak position, $\omega_0(J_\perp)$, whereas the absolute energies of higher resonances $\omega_n(J_\perp)$ with $n>1$ evolve smoothly -- as confirmed by a plot over $\omega$ instead of $\Delta \omega$, see \cref{supp:CTKS}. The appearance of spectral weight transfer and level repulsion is the hallmark of hybridization between two low-lying pairing channels.

\section{Effective two-channel model}
\label{sec:model}
Next we describe how the effective single-channel model of the doped \gls{AFM} can be extended by a second pairing channel, in order to explain and interpret the numerically observed level crossing. To this end we include a spin-singlet pair of two independent magnetic polarons, which can interact and recombine~\cite{Homeier2025} into the tightly bound bipolaronic pair that we found to dominate the two-hole spectrum in the Ising limit in \cref{fig:jperp-scan} a)-b). 

The two-channel model we employ can be motivated by the meson picture of doped \glspl{AFM}, which describes their elementary excitations as arising from two constituting partons: a spinon (s) and a chargon (c). In the regime with long-range \gls{AFM} order considered here, partons are confined and do not exist as individual, fractionalized quasiparticle excitations. Instead, the low-energy excitations are mesonic composites with an internal parton structure: (sc) correspond to magnetic polarons~\cite{Beran1996,Grusdt2018,Grusdt2019PRB,Bohrdt2020,Bohrdt2021}; (cc) correspond to tightly-bound bipolaronic hole pairs~\cite{Shraiman1988b,Vidmar2013,Grusdt2023,Bohrdt2023}; and (ss) represent spin-$1$ magnon excitations~\cite{Piazza2015}. 

The inclusion of the (sc) channel in the effective model is sufficient to explain the emergence of an additional branch in the two-hole spectrum per-se, as observed in \cref{fig:jperp-scan} c). In order to also capture the avoided level crossing between the branches, revealed in \cref{fig:jp-scan}, the following inter-channel scattering process is of central importance:
\begin{equation}
    (\textrm{cc}) \leftrightarrow (\textrm{sc})_\uparrow + (\textrm{sc})_\downarrow\,.
\end{equation}
Microscopically, this process corresponds to a breaking of the geometric string of displaced spins~\cite{Manousakis2007,Homeier2025} connecting two chargons in the (cc) state, driven by spin-exchange terms $\propto J_\perp$ or next-nearest neighbor tunneling $t'$ (not taken into account in this article)~\cite{Homeier2024,Homeier2025}.

\emph{Effective model.--}
The effective two-channel model of coupled (sc) and (cc) branches can be described by the Hamiltonian
\begin{equation}\label{eq:eff-model}
    \hat{\mathcal{H}} = \hat{\mathcal{H}}_0^{(\textrm{sc})} + \hat{\mathcal{H}}_0^{(\textrm{cc})} + \hat{\mathcal{H}}_{\textrm{int}}\,,
\end{equation}
with
\begin{align}
    \hat{\mathcal{H}}_0^{(\textrm{sc})} &= \sum_{\mathbf{k}, \sigma, n} \epsilon^{(\textrm{sc})}_{\mathbf{k},n}\hat{f}_{\mathbf{k},\sigma,n}^{\dagger}\hat{f}_{\mathbf{k},\sigma,n}\\
    \hat{\mathcal{H}}_0^{(\textrm{cc})} &= \sum_{\mathbf{k}} \epsilon^{(\textrm{cc})}_{\mathbf{k}}\hat{b}_{\mathbf{k}}^{\dagger}\hat{b}_{\mathbf{k}}\\
    \hat{\mathcal{H}}_{\textrm{int}} &= \sum_{\substack{\mathbf{k}, \mathbf{p}\\nn'}} \left( V_{\mathbf{k},\mathbf{p}}^{nn'} \hat{f}_{\mathbf{k},\uparrow,n}^{\dagger}\hat{f}_{\mathbf{p},\downarrow,n'}^{\dagger}\hat{b}_{\mathbf{k}+\mathbf{p}} + \textrm{H.c.}\right) \,.
\end{align}
Here $\hat{f}_{\mathbf{k},\sigma,n}^{\dagger}$ and $\hat{b}_{\mathbf{k}}^{\dagger}$ create a (sc) and (cc) respectively, with momentum $\mathbf{k}$ and spin $\sigma = \,\uparrow, \downarrow$. We include the first $n=1...10$ internal ro-vibrational excitations of the (sc)~\cite{Simons1990,Grusdt2018,Bohrdt2021,Bermes2024} but focus on the lowest (cc) band (with $d$-wave character)~\cite{Grusdt2023,Homeier2024}. $V_{\mathbf{k},\mathbf{p}}^{nn'}$ denotes the interaction vertex between both channels, and $\epsilon^{(\textrm{sc})}_{\mathbf{k},n}$ and $\epsilon^{(\textrm{cc})}_{\mathbf{k}}$ are the free dispersion relations of (sc) and (cc) mesons. We ignore background contributions from direct scattering of magnetic polarons / (sc)s off each other that does not involve recombination into the (cc) channel.

We compute the coupling parameters, $\epsilon^{(\textrm{sc})}_{\mathbf{k},n}$ and $V_{\mathbf{k},\mathbf{p}}^{nn'}$, from a first-principles truncated basis approach to the $t-J$ model, see~\cref{app:TwoChanMdl} and Refs.~\cite{Bermes2024,Homeier2024}. However, to improve the accuracy, we parametrize the (cc) dispersion by
\begin{equation}
    \label{eq:disp-cc}
    \epsilon^{(\textrm{cc})}_{\mathbf{k}} =  A^{(\textrm{cc})}\,\left[\textrm{cos}(k_x)  + \textrm{cos}(k_y) + 2\right] + \Delta E\,,
\end{equation}
with $A^{(\textrm{cc})}=0.66\,J$ determined by comparison to \gls{MPS} simulations~\cite{Bohrdt2023} for $t/J = 3$. With $A^{(\textrm{cc})}=0.66\,J$ fixed, we introduce the only remaining free fitting parameter of our model, the bare energy offset $\Delta E = \textrm{min}_\mathbf{k}(\epsilon^{(\textrm{cc})}_{\mathbf{k}}) - 2\,\textrm{min}_\mathbf{k}(\epsilon^{(\textrm{sc})}_{\mathbf{k}})$. Although $\Delta E$ can be estimated from our first-principles calculations, these are not accurate enough to reach the required level of precision; for example, dressing by spin-wave excitations is neglected~\cite{Bermes2024}. On the other hand, the location of the scattering resonance (i.e., the avoided level crossing) at $\Delta E = 0$ depends sensitively on the energy off-set between the two channels. The properties of the resonance itself are more universal, however, and expected to be insensitive to this single free fitting parameter within our model.  

\emph{Emergent Feshbach resonance.--}
We compare our diagrammatic calculation of the (cc) spectrum including corrections from (sc), see \cref{app:cc-self-energy}, to our high-resolution \gls{MPS} simulations. By matching the relative heights of the two lowest peaks in Fig.~\ref{fig:jp-scan} a)-b), we fit $\Delta E(J_\perp)$. This allows us to estimate the location of the scattering resonance on the $4$-leg cylinder, 
\begin{equation}
    (J_\perp / J_z)_{\rm res} \approx 0.95(5),
\end{equation}
signaled by the divergence of $1/\Delta E \to \infty$, see~\cref{fig:jp-scan} c). 
Furthermore, to characterize the width of the resonance, we compute the ratio $|V|/\Delta E$. Here $|V|$ is the strength of the (sc)-(cc) coupling of the lowest band, $|V_{\mathbf{k},\mathbf{p}}^{11}|$, averaged over $\mathbf{k}$ and $\mathbf{p}$, which is directly related to the level repulsion observed numerically in \cref{fig:jperp-scan} c). From~\cref{fig:jp-scan} c) we conclude that the $SU(2)$ invariant $t-J$ model with $J_\perp = J_z$ -- most relevant to cuprate superconductors -- is still well within the strong-coupling regime, $|V| > 2 \Delta E$, suggesting that magnetic polarons experience near-resonant $d$-wave interactions.

\begin{figure}
    \centering  \includegraphics{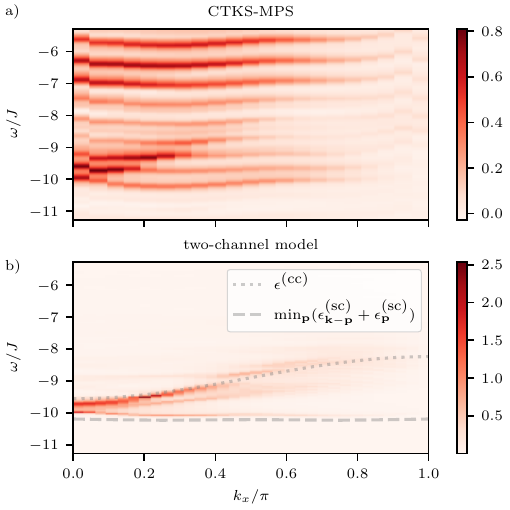}
    \caption{Comparison of the d-wave ($m_4 = 2$) two-hole spectrum $A^{(2)}(\mathbf{k},\omega)$ computed numerically, a), to the effective two-channel model, b). We study the $SU(2)$ invariant $t-J$ model ($J_\perp = J_z=J$) at $k_y = \pi/2$ as a function of $k_x$. In a) we performed high-resolution \gls{CTKS}-\gls{MPS} simulations on a $40 \times 4$ cylinder. For the two-channel model calculations in b) we used the fit parameter $\Delta E = 0.15 J$. We indicate the lower edge of the two-hole scattering continuum (dashed) and the dispersion of the tightly-bound (cc) pair (dotted).}    \label{fig:comp_spectra_sebastian}
\end{figure}

\emph{Two-channel two-hole spectrum.--}
In Fig.~\ref{fig:comp_spectra_sebastian} we compare the low-energy parts of the two-hole spectrum with $d$-wave symmetry ($m_4 = 2$) obtained from high-resolution \gls{MPS} and the effective two-channel model~\eqref{eq:eff-model}. We consider the $SU(2)$ invariant $t-J$ model, where our fitting procedure for $\Delta E$ described above yields strong coupling of the (sc) and (cc) channels. To enable a direct comparison, we solved the effective model on a $64\times 4$ cylinder similar to the $40\times4$ cylinder used for our \gls{MPS} simulations, leading to the same finite-size features associated with the short cylinder width (see \cref{app:2d-sys} for computations in an extended $32 \times 32$ system, yielding qualitatively similar results). 

We find that the effective two-channel model captures the qualitative features of the full \gls{MPS} simulations remarkably well. In particular, the same number of finite-size peaks is found, separated by small avoided crossings due to the coupling of the (cc) state to the (sc) scattering continuum. As shown in \cref{fig:comp_spectra_sebastian}, the general structure of the low-energy two-hole spectrum can be understood from the bare dispersions of the (cc), $\epsilon^{(\textrm{cc})}(\mathbf{k})$, and the lower edge of the (sc) scattering continuum, ${\rm min}_\mathbf{p} ( \epsilon^{(\textrm{sc})}_\mathbf{k-p} +\epsilon^{(\textrm{sc})}_\mathbf{p})$. Using the effective two-channel model, we confirmed that the hybridization of (cc) and (sc) branches -- the central finding from our large-scale \gls{MPS} simulations -- is a direct consequence of the inter-channel coupling $V_{\mathbf{k},\mathbf{p}}$.

\section{Outlook and Conclusions}
\label{sec:conclusion}
Measuring two-hole spectra constitutes an experimental challenge. In cuprates, coincidence and correlation \gls{ARPES}~\cite{Berakdar1998,Su2020,Mahmood2022,Kemper2025} can provide access to momentum-resolved two-electron spectra, but requires further advances in the achievable signal to noise ratio. Here we propose to use cold atom quantum simulators for a direct measurement of the two-hole spectrum, in a clean realization of a doped Hubbard model with tunable parameters~\cite{Tarruell2018,Gross2017,Bohrdt2021coldatoms}. As we show in the Methods Sec.~\ref{secMethods}, by realizing an attractive Hubbard model and probing the linear response to a Raman assisted tunneling pulse that involves a spin-flip process, the desired two-hole spectrum of the corresponding repulsive Hubbard model~\cite{Ho2009} can be directly accessed. In particular, we checked (see \cref{supp:spectroscopy}) that the predicted avoided level crossing remains visible in the Raman spectrum despite a mixing of $C_4$ angular momentum channels in this probe.

Beyond direct experimental checks of our theoretical predictions in a Hubbard model, an implementation of two-hole spectroscopy in a quantum simulator allows to test the effective two-channel picture beyond the regimes accessible to the state-of-the-art numerics presented here. On one hand, computational resources restrict our numerically exact part of our analysis to narrow four-leg cylinders, whereas cold atom experiments can be used to measure two-hole spectra in extended two-dimensional (2D) systems. Previous theoretical simulations of two-hole ground states revealing two distinct types of paired states, in Hubbard and $t-J$ models and on wider $6$-leg cylinders~\cite{Blatz2025}, suggest that the predicted avoided level crossing persists in large 2D Hubbard settings.  

On the other hand, our calculations in this article focus on the ultra-low doping regime, where the absence of competing ordered phases -- such as stripes~\cite{Corboz2014,Qin2020} -- avoids further numerical obstacles and provides a clean perspective on the properties of hole pairs. Performing cold atom experiments, one can directly measure two-hole spectra at larger dopings and low temperatures, but before ordered stripe-phases are realized~\cite{Xu2025a}. This paves the way for an experimental test of the two-channel model in the most interesting intermediate doping regime. Combining recent advances of neural quantum states for strongly correlated fermions~\cite{roth2025arXiv,Lange2024} and in their time-evolution methods~\cite{VandeWalle2025}, studies of two-particle spectra in larger 2D systems at finite doping may also become feasible.

To conclude, using ultra-high resolution numerical spectroscopy we have discovered signatures of two distinct but strongly coupled pairing channels in the $SU(2)$-invariant $t-J$ model relevant for high-$T_c$ cuprate superconductors. Our findings can be interpreted in terms of an emergent Feshbach resonance~\cite{Homeier2025}, arising from generic confinement physics in doped \glspl{AFM} and distinct from conventional atomic Feshbach resonances. The associated near-resonant interactions we expect between magnetic polarons shed new light on the non-\gls{BCS} characteristics of underdoped cuprates, and call for future investigations at finite doping, e.g. in ultracold atom experiments. Such studies will also pave the way for future refinements of the effective two-channel model, e.g. to include collective spin-wave excitations.

% % % % % % % % % % % % % % % % % % % % % % % % 
\section{Methods}
\label{secMethods}
% % % % % % % % % % % % % % % % % % % % % % % % 

\subsection{Complex-time Krylov space expansion}
\label{sec:CTKS}
The high-resolution simulations we perform in this article are based on a completely independent implementation of time-dependent \gls{MPS} simulations from our earlier work in Ref.~\cite{Bohrdt2023}, providing a valuable benchmark. In order to compute the two-particle Green's function, Eq.~\eqref{eqDefG}, with ultra-high accuracy, we employ an approach put forward in Ref.~\cite{Paeckel2024}, based on augmenting the time-evolution of the excited two-hole state using a~\gls{CTKS}. The central idea is to decompose the time integration required to obtain the frequency-resolved spectrum via
\begin{align}
    A^{(m_4)}(\mathbf{k},\omega)
    &=
    -\mathrm i
    \left[\int_0^T + \int_T^{2T} + \cdots\right]
    \mathrm dt\,
    \langle\varphi^{m_4}_\mathbf{k}\vert \hat U(t,\omega)\vert \varphi^{m_4}_\mathbf{k}\rangle \notag\\
    &=
    -\mathrm i \int_0^T \mathrm dt\,
    \langle\varphi^{m_4}_\mathbf{k}\vert \hat U(t,\omega) \hat S(T,\omega)\vert\varphi^{m_4}_\mathbf{k}\rangle \;, 
    \label{eq:spectral-function:ctks}
\end{align}
where $\hat U(t,\omega) = \mathrm e^{-\mathrm i(\hat{\mathcal{H}} - E_0 - \omega)t}$ and the initial two-hole excitation is given by $\ket{\varphi^{m_4}_\mathbf{k}} \propto \sum_{\mathbf{j}} e^{- i \mathbf{k} \cdot \mathbf{j}} |\varphi^{m_4}_\mathbf{j} \rangle $. Moreover, $\hat S(T,\omega) = \sum_{p=0}^\infty \left[\hat U(T,\omega)\right]^p$ is the boost operator, which extends the integral over the finite-time domain $[0,T)$ to the infinite-time limit. Approximating $\hat S(T,\omega)$ in a complex-time Krylov space,~\cref{eq:spectral-function:ctks} allows to overcome the Nyquist-Shannon limit and achieve ultra-high frequency resolution -- while avoiding time-evolution beyond the finite interval $T$ that is limited by entanglement generation. A short survey of the \gls{CTKS} method, including technical details, is provided in~\cref{supp:CTKS}.

\subsection{Two-particle spectroscopy in cold atom quantum simulators}
\label{sec:experiment}
To realize the required two-fermion process corresponding to the application of the pair operator $\hat{\Delta}^{(\textrm{s})}(\mathbf{j}; m_4)$, see~\cref{eq:pair-creation-operator}, we leverage the exact mapping from the repulsive Hubbard model of interest to an attractive Hubbard model with on-site interaction $\tilde{U} = -U <0$ \cite{Micnas1990, Moreo2007, Ho2009}. For a particle-hole symmetric system, this mapping corresponds to a particle-hole transformation of one spin ($\downarrow$) only,
\begin{align}
\label{eq:mapping}
    \hat{c}_{\mathbf{j}, \downarrow} &\mapsto  (-1)^{j_x+j_y}\,\hat{c}_{\mathbf{j}, \downarrow}^{\dagger}\,,\\
    \hat{c}_{\mathbf{j}, \uparrow} &\mapsto  \hat{c}_{\mathbf{j}, \uparrow}\,,
\end{align}
which maps pair creation to a spin-orbit process,
\begin{equation}
\label{eq:pair-creation-operator-attr}
    \hat{\tilde{\Delta}}^{(\textrm{s})}(\mathbf{j}; m_4) = (-1)^{j_x+j_y}\sum_{\mathbf{i}:\langle\mathbf{i},\mathbf{j}\rangle}e^{im_4\varphi_{\mathbf{i} - \mathbf{j}}}(\hat{c}_{\mathbf{j},\downarrow}^\dagger\hat{c}_{\mathbf{i},\uparrow} - \hat{c}_{\mathbf{i},\downarrow}^\dagger\hat{c}_{\mathbf{j},\uparrow})\,.
\end{equation}
I.e., by experimentally realizing an attractive Hubbard model~\cite{Mitra2018, Gall2020, Chan2020} and measuring dynamical correlations of the form $\langle \hat{\tilde{\Delta}}^\dagger(t) \hat{\tilde{\Delta}}(0) \rangle$, direct access is obtained to the two-hole Green's function $\mathcal{G}(t)$ of the corresponding repulsive Hubbard model. 

Experimentally, the process~\cref{eq:pair-creation-operator-attr} consisting of a spin-flip combined with nearest-neighbor hopping, can be realized by Raman assisted tunneling~\cite{Jaksch2003,Aidelsburger2011}. Two running Raman lasers with wave vectors $\mathbf{k}_1$ and $\mathbf{k}_2$ in a $\Lambda$ configuration create excitations with momentum $\hbar \mathbf{k} = \hbar(\mathbf{k}_1 - \mathbf{k}_2)$ and energy $\hbar \omega = \hbar c (|\mathbf{k}_1| - |\mathbf{k}_2|)$; see~\cref{supp:spectroscopy} for more details. 
A main advantage of the proposed Raman scheme is the full tunability of the total momentum transfer $\mathbf{k}$, which can be tuned by changing the angle between the two laser beams. This corresponds directly to the momentum of the target excitation, which is shifted by $(\pi, \pi)$ when transforming back to the corresponding repulsive Hubbard model using Eq.~\eqref{eq:mapping}; i.e., a pair excitation with momentum $\mathbf{k}_{\textrm{rep}}$ in the repulsive Hubbard model corresponds to an excitation with momentum $\mathbf{k}_{\textrm{attr}} = \mathbf{k}_{\textrm{rep}} - (\pi,\pi)$ probed directly in the attractive model.

Since the two Raman beams break the discrete lattice-rotational symmetry, $C_4$ angular momentum cannot be directly controlled in this scheme. Instead, a superposition of different $m_4$ components of the pair spectrum $A^{(m_4)}(\mathbf{k},\omega)$ is probed, with relative weights depending on the incident momenta $\mathbf{k}_{1,2}$. In \cref{supp:spectroscopy} we explicitly compute the relevant matrix elements and demonstrate that this is sufficient to probe the avoided level crossing that we predict in the $d$-wave channel.

\section*{Acknowledgments}
We thank Eugen Dizer, Eugene Demler and  Richard Schmidt for fruitful discussions.
--
This project has received funding from the European Research Council (ERC) under the European Union’s Horizon 2020 research and innovation programm (Grant Agreement no 948141) — ERC Starting Grant SimUcQuam, and from the Deutsche Forschungsgemeinschaft (DFG, German Research Foundation) under Germany's Excellence Strategy -- EXC-2111 -- 390814868.
LH acknowledges support by the Simons Collaboration on Ultra-Quantum Matter, which is a grant from the Simons Foundation (651440), and by Studienstiftung des deutschen Volkes.

\bibliographystyle{apsrev4-2}

\begin{thebibliography}{75}%
\makeatletter
\providecommand \@ifxundefined [1]{%
 \@ifx{#1\undefined}
}%
\providecommand \@ifnum [1]{%
 \ifnum #1\expandafter \@firstoftwo
 \else \expandafter \@secondoftwo
 \fi
}%
\providecommand \@ifx [1]{%
 \ifx #1\expandafter \@firstoftwo
 \else \expandafter \@secondoftwo
 \fi
}%
\providecommand \natexlab [1]{#1}%
\providecommand \enquote  [1]{``#1''}%
\providecommand \bibnamefont  [1]{#1}%
\providecommand \bibfnamefont [1]{#1}%
\providecommand \citenamefont [1]{#1}%
\providecommand \href@noop [0]{\@secondoftwo}%
\providecommand \href [0]{\begingroup \@sanitize@url \@href}%
\providecommand \@href[1]{\@@startlink{#1}\@@href}%
\providecommand \@@href[1]{\endgroup#1\@@endlink}%
\providecommand \@sanitize@url [0]{\catcode `\\12\catcode `\$12\catcode
  `\&12\catcode `\#12\catcode `\^12\catcode `\_12\catcode `\%12\relax}%
\providecommand \@@startlink[1]{}%
\providecommand \@@endlink[0]{}%
\providecommand \url  [0]{\begingroup\@sanitize@url \@url }%
\providecommand \@url [1]{\endgroup\@href {#1}{\urlprefix }}%
\providecommand \urlprefix  [0]{URL }%
\providecommand \Eprint [0]{\href }%
\providecommand \doibase [0]{https://doi.org/}%
\providecommand \selectlanguage [0]{\@gobble}%
\providecommand \bibinfo  [0]{\@secondoftwo}%
\providecommand \bibfield  [0]{\@secondoftwo}%
\providecommand \translation [1]{[#1]}%
\providecommand \BibitemOpen [0]{}%
\providecommand \bibitemStop [0]{}%
\providecommand \bibitemNoStop [0]{.\EOS\space}%
\providecommand \EOS [0]{\spacefactor3000\relax}%
\providecommand \BibitemShut  [1]{\csname bibitem#1\endcsname}%
\let\auto@bib@innerbib\@empty
%</preamble>
\bibitem [{\citenamefont {Lee}\ \emph {et~al.}(2006)\citenamefont {Lee},
  \citenamefont {Nagaosa},\ and\ \citenamefont {Wen}}]{Lee2006}%
  \BibitemOpen
  \bibfield  {author} {\bibinfo {author} {\bibfnamefont {P.~A.}\ \bibnamefont
  {Lee}}, \bibinfo {author} {\bibfnamefont {N.}~\bibnamefont {Nagaosa}},\ and\
  \bibinfo {author} {\bibfnamefont {X.-G.}\ \bibnamefont {Wen}},\ }\href
  {https://doi.org/10.1103/RevModPhys.78.17} {\bibfield  {journal} {\bibinfo
  {journal} {Rev. Mod. Phys.}\ }\textbf {\bibinfo {volume} {78}},\ \bibinfo
  {pages} {17} (\bibinfo {year} {2006})}\BibitemShut {NoStop}%
\bibitem [{\citenamefont {Bednorz}\ and\ \citenamefont
  {M{\"u}ller}(1986)}]{Bednorz1986}%
  \BibitemOpen
  \bibfield  {author} {\bibinfo {author} {\bibfnamefont {J.~G.}\ \bibnamefont
  {Bednorz}}\ and\ \bibinfo {author} {\bibfnamefont {K.~A.}\ \bibnamefont
  {M{\"u}ller}},\ }\href@noop {} {\bibfield  {journal} {\bibinfo  {journal}
  {Zeitschrift f{\"u}r Physik B Condensed Matter}\ }\textbf {\bibinfo {volume}
  {64}},\ \bibinfo {pages} {189} (\bibinfo {year} {1986})}\BibitemShut
  {NoStop}%
\bibitem [{\citenamefont {Uemura}\ \emph {et~al.}(1991)\citenamefont {Uemura},
  \citenamefont {Le}, \citenamefont {Luke}, \citenamefont {Sternlieb},
  \citenamefont {Wu}, \citenamefont {Brewer}, \citenamefont {Riseman},
  \citenamefont {Seaman}, \citenamefont {Maple}, \citenamefont {Ishikawa},
  \citenamefont {Hinks}, \citenamefont {Jorgensen}, \citenamefont {Saito},\
  and\ \citenamefont {Yamochi}}]{Uemura1991}%
  \BibitemOpen
  \bibfield  {author} {\bibinfo {author} {\bibfnamefont {Y.~J.}\ \bibnamefont
  {Uemura}}, \bibinfo {author} {\bibfnamefont {L.~P.}\ \bibnamefont {Le}},
  \bibinfo {author} {\bibfnamefont {G.~M.}\ \bibnamefont {Luke}}, \bibinfo
  {author} {\bibfnamefont {B.~J.}\ \bibnamefont {Sternlieb}}, \bibinfo {author}
  {\bibfnamefont {W.~D.}\ \bibnamefont {Wu}}, \bibinfo {author} {\bibfnamefont
  {J.~H.}\ \bibnamefont {Brewer}}, \bibinfo {author} {\bibfnamefont {T.~M.}\
  \bibnamefont {Riseman}}, \bibinfo {author} {\bibfnamefont {C.~L.}\
  \bibnamefont {Seaman}}, \bibinfo {author} {\bibfnamefont {M.~B.}\
  \bibnamefont {Maple}}, \bibinfo {author} {\bibfnamefont {M.}~\bibnamefont
  {Ishikawa}}, \bibinfo {author} {\bibfnamefont {D.~G.}\ \bibnamefont {Hinks}},
  \bibinfo {author} {\bibfnamefont {J.~D.}\ \bibnamefont {Jorgensen}}, \bibinfo
  {author} {\bibfnamefont {G.}~\bibnamefont {Saito}},\ and\ \bibinfo {author}
  {\bibfnamefont {H.}~\bibnamefont {Yamochi}},\ }\href
  {https://doi.org/10.1103/PhysRevLett.66.2665} {\bibfield  {journal} {\bibinfo
   {journal} {Phys. Rev. Lett.}\ }\textbf {\bibinfo {volume} {66}},\ \bibinfo
  {pages} {2665} (\bibinfo {year} {1991})}\BibitemShut {NoStop}%
\bibitem [{\citenamefont {Emery}\ and\ \citenamefont
  {Kivelson}(1995)}]{Emery1995}%
  \BibitemOpen
  \bibfield  {author} {\bibinfo {author} {\bibfnamefont {V.~J.}\ \bibnamefont
  {Emery}}\ and\ \bibinfo {author} {\bibfnamefont {S.~A.}\ \bibnamefont
  {Kivelson}},\ }\href {https://doi.org/10.1038/374434a0} {\bibfield  {journal}
  {\bibinfo  {journal} {Nature}\ }\textbf {\bibinfo {volume} {374}},\ \bibinfo
  {pages} {434} (\bibinfo {year} {1995})}\BibitemShut {NoStop}%
\bibitem [{\citenamefont {Niu}\ \emph {et~al.}(2024)\citenamefont {Niu},
  \citenamefont {Larrazabal}, \citenamefont {Gozlinski}, \citenamefont {Sato},
  \citenamefont {Bastiaans}, \citenamefont {Benschop}, \citenamefont {Ge},
  \citenamefont {Blanter}, \citenamefont {Gu}, \citenamefont {Swart},\ and\
  \citenamefont {Allan}}]{niu2024arXiv}%
  \BibitemOpen
  \bibfield  {author} {\bibinfo {author} {\bibfnamefont {J.}~\bibnamefont
  {Niu}}, \bibinfo {author} {\bibfnamefont {M.~O.}\ \bibnamefont {Larrazabal}},
  \bibinfo {author} {\bibfnamefont {T.}~\bibnamefont {Gozlinski}}, \bibinfo
  {author} {\bibfnamefont {Y.}~\bibnamefont {Sato}}, \bibinfo {author}
  {\bibfnamefont {K.~M.}\ \bibnamefont {Bastiaans}}, \bibinfo {author}
  {\bibfnamefont {T.}~\bibnamefont {Benschop}}, \bibinfo {author}
  {\bibfnamefont {J.-F.}\ \bibnamefont {Ge}}, \bibinfo {author} {\bibfnamefont
  {Y.~M.}\ \bibnamefont {Blanter}}, \bibinfo {author} {\bibfnamefont
  {G.}~\bibnamefont {Gu}}, \bibinfo {author} {\bibfnamefont {I.}~\bibnamefont
  {Swart}},\ and\ \bibinfo {author} {\bibfnamefont {M.~P.}\ \bibnamefont
  {Allan}},\ }\href {https://arxiv.org/abs/2409.15928} {\bibfield  {journal}
  {\bibinfo  {journal} {arXiv:2409.15928}\ } (\bibinfo {year}
  {2024})}\BibitemShut {NoStop}%
\bibitem [{\citenamefont {Wollman}\ \emph {et~al.}(1993)\citenamefont
  {Wollman}, \citenamefont {Van~Harlingen}, \citenamefont {Lee}, \citenamefont
  {Ginsberg},\ and\ \citenamefont {Leggett}}]{Wollman1993}%
  \BibitemOpen
  \bibfield  {author} {\bibinfo {author} {\bibfnamefont {D.~A.}\ \bibnamefont
  {Wollman}}, \bibinfo {author} {\bibfnamefont {D.~J.}\ \bibnamefont
  {Van~Harlingen}}, \bibinfo {author} {\bibfnamefont {W.~C.}\ \bibnamefont
  {Lee}}, \bibinfo {author} {\bibfnamefont {D.~M.}\ \bibnamefont {Ginsberg}},\
  and\ \bibinfo {author} {\bibfnamefont {A.~J.}\ \bibnamefont {Leggett}},\
  }\href {https://doi.org/10.1103/PhysRevLett.71.2134} {\bibfield  {journal}
  {\bibinfo  {journal} {Phys. Rev. Lett.}\ }\textbf {\bibinfo {volume} {71}},\
  \bibinfo {pages} {2134} (\bibinfo {year} {1993})}\BibitemShut {NoStop}%
\bibitem [{\citenamefont {Hashimoto}\ \emph {et~al.}(2014)\citenamefont
  {Hashimoto}, \citenamefont {Vishik}, \citenamefont {He}, \citenamefont
  {Devereaux},\ and\ \citenamefont {Shen}}]{Hashimoto2014}%
  \BibitemOpen
  \bibfield  {author} {\bibinfo {author} {\bibfnamefont {M.}~\bibnamefont
  {Hashimoto}}, \bibinfo {author} {\bibfnamefont {I.~M.}\ \bibnamefont
  {Vishik}}, \bibinfo {author} {\bibfnamefont {R.-H.}\ \bibnamefont {He}},
  \bibinfo {author} {\bibfnamefont {T.~P.}\ \bibnamefont {Devereaux}},\ and\
  \bibinfo {author} {\bibfnamefont {Z.-X.}\ \bibnamefont {Shen}},\ }\href
  {https://doi.org/10.1038/nphys3009} {\bibfield  {journal} {\bibinfo
  {journal} {Nature Phys}\ }\textbf {\bibinfo {volume} {10}},\ \bibinfo {pages}
  {483} (\bibinfo {year} {2014})}\BibitemShut {NoStop}%
\bibitem [{\citenamefont {Dagotto}\ \emph {et~al.}(1990)\citenamefont
  {Dagotto}, \citenamefont {Riera},\ and\ \citenamefont
  {Young}}]{Dagotto1990pair}%
  \BibitemOpen
  \bibfield  {author} {\bibinfo {author} {\bibfnamefont {E.}~\bibnamefont
  {Dagotto}}, \bibinfo {author} {\bibfnamefont {J.}~\bibnamefont {Riera}},\
  and\ \bibinfo {author} {\bibfnamefont {A.~P.}\ \bibnamefont {Young}},\ }\href
  {https://doi.org/10.1103/PhysRevB.42.2347} {\bibfield  {journal} {\bibinfo
  {journal} {Phys. Rev. B}\ }\textbf {\bibinfo {volume} {42}},\ \bibinfo
  {pages} {2347} (\bibinfo {year} {1990})}\BibitemShut {NoStop}%
\bibitem [{\citenamefont {Corboz}\ \emph {et~al.}(2014)\citenamefont {Corboz},
  \citenamefont {Rice},\ and\ \citenamefont {Troyer}}]{Corboz2014}%
  \BibitemOpen
  \bibfield  {author} {\bibinfo {author} {\bibfnamefont {P.}~\bibnamefont
  {Corboz}}, \bibinfo {author} {\bibfnamefont {T.~M.}\ \bibnamefont {Rice}},\
  and\ \bibinfo {author} {\bibfnamefont {M.}~\bibnamefont {Troyer}},\ }\href
  {https://doi.org/10.1103/PhysRevLett.113.046402} {\bibfield  {journal}
  {\bibinfo  {journal} {Phys. Rev. Lett.}\ }\textbf {\bibinfo {volume} {113}},\
  \bibinfo {pages} {046402} (\bibinfo {year} {2014})}\BibitemShut {NoStop}%
\bibitem [{\citenamefont {Xu}\ \emph {et~al.}(2024)\citenamefont {Xu},
  \citenamefont {Chung}, \citenamefont {Qin}, \citenamefont {Schollw{\"o}ck},
  \citenamefont {White},\ and\ \citenamefont {Zhang}}]{Xu2024}%
  \BibitemOpen
  \bibfield  {author} {\bibinfo {author} {\bibfnamefont {H.}~\bibnamefont
  {Xu}}, \bibinfo {author} {\bibfnamefont {C.-M.}\ \bibnamefont {Chung}},
  \bibinfo {author} {\bibfnamefont {M.}~\bibnamefont {Qin}}, \bibinfo {author}
  {\bibfnamefont {U.}~\bibnamefont {Schollw{\"o}ck}}, \bibinfo {author}
  {\bibfnamefont {S.~R.}\ \bibnamefont {White}},\ and\ \bibinfo {author}
  {\bibfnamefont {S.}~\bibnamefont {Zhang}},\ }\href
  {https://doi.org/10.1126/science.adh7691} {\bibfield  {journal} {\bibinfo
  {journal} {Science}\ }\textbf {\bibinfo {volume} {384}},\ \bibinfo {pages}
  {eadh7691} (\bibinfo {year} {2024})}\BibitemShut {NoStop}%
\bibitem [{\citenamefont {Scalapino}\ \emph {et~al.}(1987)\citenamefont
  {Scalapino}, \citenamefont {Loh},\ and\ \citenamefont
  {Hirsch}}]{Scalapino1987}%
  \BibitemOpen
  \bibfield  {author} {\bibinfo {author} {\bibfnamefont {D.~J.}\ \bibnamefont
  {Scalapino}}, \bibinfo {author} {\bibfnamefont {E.}~\bibnamefont {Loh}},\
  and\ \bibinfo {author} {\bibfnamefont {J.~E.}\ \bibnamefont {Hirsch}},\
  }\href {https://doi.org/10.1103/PhysRevB.35.6694} {\bibfield  {journal}
  {\bibinfo  {journal} {Phys. Rev. B}\ }\textbf {\bibinfo {volume} {35}},\
  \bibinfo {pages} {6694} (\bibinfo {year} {1987})}\BibitemShut {NoStop}%
\bibitem [{\citenamefont {Halboth}\ and\ \citenamefont
  {Metzner}(2000)}]{Halboth2000}%
  \BibitemOpen
  \bibfield  {author} {\bibinfo {author} {\bibfnamefont {C.~J.}\ \bibnamefont
  {Halboth}}\ and\ \bibinfo {author} {\bibfnamefont {W.}~\bibnamefont
  {Metzner}},\ }\href {https://doi.org/10.1103/PhysRevLett.85.5162} {\bibfield
  {journal} {\bibinfo  {journal} {Phys. Rev. Lett.}\ }\textbf {\bibinfo
  {volume} {85}},\ \bibinfo {pages} {5162} (\bibinfo {year}
  {2000})}\BibitemShut {NoStop}%
\bibitem [{\citenamefont {Metzner}\ \emph {et~al.}(2012)\citenamefont
  {Metzner}, \citenamefont {Salmhofer}, \citenamefont {Honerkamp},
  \citenamefont {Meden},\ and\ \citenamefont {Sch{\"o}nhammer}}]{Metzner2012}%
  \BibitemOpen
  \bibfield  {author} {\bibinfo {author} {\bibfnamefont {W.}~\bibnamefont
  {Metzner}}, \bibinfo {author} {\bibfnamefont {M.}~\bibnamefont {Salmhofer}},
  \bibinfo {author} {\bibfnamefont {C.}~\bibnamefont {Honerkamp}}, \bibinfo
  {author} {\bibfnamefont {V.}~\bibnamefont {Meden}},\ and\ \bibinfo {author}
  {\bibfnamefont {K.}~\bibnamefont {Sch{\"o}nhammer}},\ }\href
  {https://doi.org/10.1103/RevModPhys.84.299} {\bibfield  {journal} {\bibinfo
  {journal} {Rev. Mod. Phys.}\ }\textbf {\bibinfo {volume} {84}},\ \bibinfo
  {pages} {299} (\bibinfo {year} {2012})}\BibitemShut {NoStop}%
\bibitem [{\citenamefont {Vilardi}\ \emph {et~al.}(2019)\citenamefont
  {Vilardi}, \citenamefont {Taranto},\ and\ \citenamefont
  {Metzner}}]{Vilardi2019}%
  \BibitemOpen
  \bibfield  {author} {\bibinfo {author} {\bibfnamefont {D.}~\bibnamefont
  {Vilardi}}, \bibinfo {author} {\bibfnamefont {C.}~\bibnamefont {Taranto}},\
  and\ \bibinfo {author} {\bibfnamefont {W.}~\bibnamefont {Metzner}},\ }\href
  {https://doi.org/10.1103/PhysRevB.99.104501} {\bibfield  {journal} {\bibinfo
  {journal} {Phys. Rev. B}\ }\textbf {\bibinfo {volume} {99}},\ \bibinfo
  {pages} {104501} (\bibinfo {year} {2019})}\BibitemShut {NoStop}%
\bibitem [{\citenamefont {Stajic}\ \emph {et~al.}(2003)\citenamefont {Stajic},
  \citenamefont {Iyengar}, \citenamefont {Levin}, \citenamefont {Boyce},\ and\
  \citenamefont {Lemberger}}]{Stajic2003}%
  \BibitemOpen
  \bibfield  {author} {\bibinfo {author} {\bibfnamefont {J.}~\bibnamefont
  {Stajic}}, \bibinfo {author} {\bibfnamefont {A.}~\bibnamefont {Iyengar}},
  \bibinfo {author} {\bibfnamefont {K.}~\bibnamefont {Levin}}, \bibinfo
  {author} {\bibfnamefont {B.~R.}\ \bibnamefont {Boyce}},\ and\ \bibinfo
  {author} {\bibfnamefont {T.~R.}\ \bibnamefont {Lemberger}},\ }\href
  {https://doi.org/10.1103/PhysRevB.68.024520} {\bibfield  {journal} {\bibinfo
  {journal} {Phys. Rev. B}\ }\textbf {\bibinfo {volume} {68}},\ \bibinfo
  {pages} {024520} (\bibinfo {year} {2003})}\BibitemShut {NoStop}%
\bibitem [{\citenamefont {Corson}\ \emph {et~al.}(1999)\citenamefont {Corson},
  \citenamefont {Mallozzi}, \citenamefont {Orenstein}, \citenamefont
  {Eckstein},\ and\ \citenamefont {Bozovic}}]{Corson1999}%
  \BibitemOpen
  \bibfield  {author} {\bibinfo {author} {\bibfnamefont {J.}~\bibnamefont
  {Corson}}, \bibinfo {author} {\bibfnamefont {R.}~\bibnamefont {Mallozzi}},
  \bibinfo {author} {\bibfnamefont {J.}~\bibnamefont {Orenstein}}, \bibinfo
  {author} {\bibfnamefont {J.~N.}\ \bibnamefont {Eckstein}},\ and\ \bibinfo
  {author} {\bibfnamefont {I.}~\bibnamefont {Bozovic}},\ }\href
  {https://doi.org/10.1038/18402} {\bibfield  {journal} {\bibinfo  {journal}
  {Nature}\ }\textbf {\bibinfo {volume} {398}},\ \bibinfo {pages} {221}
  (\bibinfo {year} {1999})}\BibitemShut {NoStop}%
\bibitem [{\citenamefont {Zhou}\ \emph {et~al.}(2019)\citenamefont {Zhou},
  \citenamefont {Chen}, \citenamefont {Liu}, \citenamefont {Sochnikov},
  \citenamefont {Bollinger}, \citenamefont {Han}, \citenamefont {Zhu},
  \citenamefont {He}, \citenamefont {Bo{\u z}ovi{\'c}},\ and\ \citenamefont
  {Natelson}}]{Zhou2019}%
  \BibitemOpen
  \bibfield  {author} {\bibinfo {author} {\bibfnamefont {P.}~\bibnamefont
  {Zhou}}, \bibinfo {author} {\bibfnamefont {L.}~\bibnamefont {Chen}}, \bibinfo
  {author} {\bibfnamefont {Y.}~\bibnamefont {Liu}}, \bibinfo {author}
  {\bibfnamefont {I.}~\bibnamefont {Sochnikov}}, \bibinfo {author}
  {\bibfnamefont {A.~T.}\ \bibnamefont {Bollinger}}, \bibinfo {author}
  {\bibfnamefont {M.-G.}\ \bibnamefont {Han}}, \bibinfo {author} {\bibfnamefont
  {Y.}~\bibnamefont {Zhu}}, \bibinfo {author} {\bibfnamefont {X.}~\bibnamefont
  {He}}, \bibinfo {author} {\bibfnamefont {I.}~\bibnamefont {Bo{\u
  z}ovi{\'c}}},\ and\ \bibinfo {author} {\bibfnamefont {D.}~\bibnamefont
  {Natelson}},\ }\href {https://doi.org/10.1038/s41586-019-1486-7} {\bibfield
  {journal} {\bibinfo  {journal} {Nature}\ }\textbf {\bibinfo {volume} {572}},\
  \bibinfo {pages} {493} (\bibinfo {year} {2019})}\BibitemShut {NoStop}%
\bibitem [{\citenamefont {Loram}\ \emph {et~al.}(1994)\citenamefont {Loram},
  \citenamefont {Mirza}, \citenamefont {Wade}, \citenamefont {Cooper},\ and\
  \citenamefont {Liang}}]{Loram1994}%
  \BibitemOpen
  \bibfield  {author} {\bibinfo {author} {\bibfnamefont {J.}~\bibnamefont
  {Loram}}, \bibinfo {author} {\bibfnamefont {K.}~\bibnamefont {Mirza}},
  \bibinfo {author} {\bibfnamefont {J.}~\bibnamefont {Wade}}, \bibinfo {author}
  {\bibfnamefont {J.}~\bibnamefont {Cooper}},\ and\ \bibinfo {author}
  {\bibfnamefont {W.}~\bibnamefont {Liang}},\ }\href
  {https://doi.org/https://doi.org/10.1016/0921-4534(94)91331-5} {\bibfield
  {journal} {\bibinfo  {journal} {Physica C: Superconductivity}\ }\textbf
  {\bibinfo {volume} {235-240}},\ \bibinfo {pages} {134} (\bibinfo {year}
  {1994})}\BibitemShut {NoStop}%
\bibitem [{\citenamefont {Sonier}\ \emph {et~al.}(2007)\citenamefont {Sonier},
  \citenamefont {Sabok-Sayr}, \citenamefont {Callaghan}, \citenamefont
  {Kaiser}, \citenamefont {Pacradouni}, \citenamefont {Brewer}, \citenamefont
  {Stubbs}, \citenamefont {Hardy}, \citenamefont {Bonn}, \citenamefont
  {Liang},\ and\ \citenamefont {Atkinson}}]{Sonier2007}%
  \BibitemOpen
  \bibfield  {author} {\bibinfo {author} {\bibfnamefont {J.~E.}\ \bibnamefont
  {Sonier}}, \bibinfo {author} {\bibfnamefont {S.~A.}\ \bibnamefont
  {Sabok-Sayr}}, \bibinfo {author} {\bibfnamefont {F.~D.}\ \bibnamefont
  {Callaghan}}, \bibinfo {author} {\bibfnamefont {C.~V.}\ \bibnamefont
  {Kaiser}}, \bibinfo {author} {\bibfnamefont {V.}~\bibnamefont {Pacradouni}},
  \bibinfo {author} {\bibfnamefont {J.~H.}\ \bibnamefont {Brewer}}, \bibinfo
  {author} {\bibfnamefont {S.~L.}\ \bibnamefont {Stubbs}}, \bibinfo {author}
  {\bibfnamefont {W.~N.}\ \bibnamefont {Hardy}}, \bibinfo {author}
  {\bibfnamefont {D.~A.}\ \bibnamefont {Bonn}}, \bibinfo {author}
  {\bibfnamefont {R.}~\bibnamefont {Liang}},\ and\ \bibinfo {author}
  {\bibfnamefont {W.~A.}\ \bibnamefont {Atkinson}},\ }\href
  {https://doi.org/10.1103/PhysRevB.76.134518} {\bibfield  {journal} {\bibinfo
  {journal} {Phys. Rev. B}\ }\textbf {\bibinfo {volume} {76}},\ \bibinfo
  {pages} {134518} (\bibinfo {year} {2007})}\BibitemShut {NoStop}%
\bibitem [{\citenamefont {Wen}\ \emph {et~al.}(2003)\citenamefont {Wen},
  \citenamefont {Yang}, \citenamefont {Li}, \citenamefont {Zeng}, \citenamefont
  {Soukiassian}, \citenamefont {Si},\ and\ \citenamefont {Xi}}]{Wen2003}%
  \BibitemOpen
  \bibfield  {author} {\bibinfo {author} {\bibfnamefont {H.~H.}\ \bibnamefont
  {Wen}}, \bibinfo {author} {\bibfnamefont {H.~P.}\ \bibnamefont {Yang}},
  \bibinfo {author} {\bibfnamefont {S.~L.}\ \bibnamefont {Li}}, \bibinfo
  {author} {\bibfnamefont {X.~H.}\ \bibnamefont {Zeng}}, \bibinfo {author}
  {\bibfnamefont {A.~A.}\ \bibnamefont {Soukiassian}}, \bibinfo {author}
  {\bibfnamefont {W.~D.}\ \bibnamefont {Si}},\ and\ \bibinfo {author}
  {\bibfnamefont {X.~X.}\ \bibnamefont {Xi}},\ }\href
  {https://doi.org/10.1209/epl/i2003-00627-1} {\bibfield  {journal} {\bibinfo
  {journal} {Europhysics Letters}\ }\textbf {\bibinfo {volume} {64}},\ \bibinfo
  {pages} {790} (\bibinfo {year} {2003})}\BibitemShut {NoStop}%
\bibitem [{\citenamefont {Sordi}\ \emph {et~al.}(2012)\citenamefont {Sordi},
  \citenamefont {Sémon}, \citenamefont {Haule},\ and\ \citenamefont
  {Tremblay}}]{Sordi2012}%
  \BibitemOpen
  \bibfield  {author} {\bibinfo {author} {\bibfnamefont {G.}~\bibnamefont
  {Sordi}}, \bibinfo {author} {\bibfnamefont {P.}~\bibnamefont {Sémon}},
  \bibinfo {author} {\bibfnamefont {K.}~\bibnamefont {Haule}},\ and\ \bibinfo
  {author} {\bibfnamefont {A.-M.~S.}\ \bibnamefont {Tremblay}},\ }\href
  {https://doi.org/10.1038/srep00547} {\bibfield  {journal} {\bibinfo
  {journal} {Scientific Reports}\ }\textbf {\bibinfo {volume} {2}},\ \bibinfo
  {pages} {547} (\bibinfo {year} {2012})}\BibitemShut {NoStop}%
\bibitem [{\citenamefont {Sch\"afer}\ \emph {et~al.}(2021)\citenamefont
  {Sch\"afer}, \citenamefont {Wentzell}, \citenamefont {\ifmmode~\check{S}\else
  \v{S}\fi{}imkovic}, \citenamefont {He}, \citenamefont {Hille}, \citenamefont
  {Klett}, \citenamefont {Eckhardt}, \citenamefont {Arzhang}, \citenamefont
  {Harkov}, \citenamefont {Le~R\'egent}, \citenamefont {Kirsch}, \citenamefont
  {Wang}, \citenamefont {Kim}, \citenamefont {Kozik}, \citenamefont {Stepanov},
  \citenamefont {Kauch}, \citenamefont {Andergassen}, \citenamefont {Hansmann},
  \citenamefont {Rohe}, \citenamefont {Vilk}, \citenamefont {LeBlanc},
  \citenamefont {Zhang}, \citenamefont {Tremblay}, \citenamefont {Ferrero},
  \citenamefont {Parcollet},\ and\ \citenamefont {Georges}}]{Schaefer2021}%
  \BibitemOpen
  \bibfield  {author} {\bibinfo {author} {\bibfnamefont {T.}~\bibnamefont
  {Sch\"afer}}, \bibinfo {author} {\bibfnamefont {N.}~\bibnamefont {Wentzell}},
  \bibinfo {author} {\bibfnamefont {F.}~\bibnamefont {\ifmmode~\check{S}\else
  \v{S}\fi{}imkovic}}, \bibinfo {author} {\bibfnamefont {Y.-Y.}\ \bibnamefont
  {He}}, \bibinfo {author} {\bibfnamefont {C.}~\bibnamefont {Hille}}, \bibinfo
  {author} {\bibfnamefont {M.}~\bibnamefont {Klett}}, \bibinfo {author}
  {\bibfnamefont {C.~J.}\ \bibnamefont {Eckhardt}}, \bibinfo {author}
  {\bibfnamefont {B.}~\bibnamefont {Arzhang}}, \bibinfo {author} {\bibfnamefont
  {V.}~\bibnamefont {Harkov}}, \bibinfo {author} {\bibfnamefont {F.~m. c.-M.}\
  \bibnamefont {Le~R\'egent}}, \bibinfo {author} {\bibfnamefont
  {A.}~\bibnamefont {Kirsch}}, \bibinfo {author} {\bibfnamefont
  {Y.}~\bibnamefont {Wang}}, \bibinfo {author} {\bibfnamefont {A.~J.}\
  \bibnamefont {Kim}}, \bibinfo {author} {\bibfnamefont {E.}~\bibnamefont
  {Kozik}}, \bibinfo {author} {\bibfnamefont {E.~A.}\ \bibnamefont {Stepanov}},
  \bibinfo {author} {\bibfnamefont {A.}~\bibnamefont {Kauch}}, \bibinfo
  {author} {\bibfnamefont {S.}~\bibnamefont {Andergassen}}, \bibinfo {author}
  {\bibfnamefont {P.}~\bibnamefont {Hansmann}}, \bibinfo {author}
  {\bibfnamefont {D.}~\bibnamefont {Rohe}}, \bibinfo {author} {\bibfnamefont
  {Y.~M.}\ \bibnamefont {Vilk}}, \bibinfo {author} {\bibfnamefont {J.~P.~F.}\
  \bibnamefont {LeBlanc}}, \bibinfo {author} {\bibfnamefont {S.}~\bibnamefont
  {Zhang}}, \bibinfo {author} {\bibfnamefont {A.-M.~S.}\ \bibnamefont
  {Tremblay}}, \bibinfo {author} {\bibfnamefont {M.}~\bibnamefont {Ferrero}},
  \bibinfo {author} {\bibfnamefont {O.}~\bibnamefont {Parcollet}},\ and\
  \bibinfo {author} {\bibfnamefont {A.}~\bibnamefont {Georges}},\ }\href
  {https://doi.org/10.1103/PhysRevX.11.011058} {\bibfield  {journal} {\bibinfo
  {journal} {Phys. Rev. X}\ }\textbf {\bibinfo {volume} {11}},\ \bibinfo
  {pages} {011058} (\bibinfo {year} {2021})}\BibitemShut {NoStop}%
\bibitem [{\citenamefont {{\v S}imkovic}\ \emph {et~al.}(2024)\citenamefont
  {{\v S}imkovic}, \citenamefont {Rossi}, \citenamefont {Georges},\ and\
  \citenamefont {Ferrero}}]{Simkovic2024}%
  \BibitemOpen
  \bibfield  {author} {\bibinfo {author} {\bibfnamefont {F.}~\bibnamefont {{\v
  S}imkovic}}, \bibinfo {author} {\bibfnamefont {R.}~\bibnamefont {Rossi}},
  \bibinfo {author} {\bibfnamefont {A.}~\bibnamefont {Georges}},\ and\ \bibinfo
  {author} {\bibfnamefont {M.}~\bibnamefont {Ferrero}},\ }\href
  {https://doi.org/10.1126/science.ade9194} {\bibfield  {journal} {\bibinfo
  {journal} {Science}\ }\textbf {\bibinfo {volume} {385}},\ \bibinfo {pages}
  {eade9194} (\bibinfo {year} {2024})}\BibitemShut {NoStop}%
\bibitem [{\citenamefont {Damascelli}\ \emph {et~al.}(2003)\citenamefont
  {Damascelli}, \citenamefont {Hussain},\ and\ \citenamefont
  {Shen}}]{Damascelli2003}%
  \BibitemOpen
  \bibfield  {author} {\bibinfo {author} {\bibfnamefont {A.}~\bibnamefont
  {Damascelli}}, \bibinfo {author} {\bibfnamefont {Z.}~\bibnamefont
  {Hussain}},\ and\ \bibinfo {author} {\bibfnamefont {Z.-X.}\ \bibnamefont
  {Shen}},\ }\href {https://doi.org/10.1103/RevModPhys.75.473} {\bibfield
  {journal} {\bibinfo  {journal} {Rev. Mod. Phys.}\ }\textbf {\bibinfo {volume}
  {75}},\ \bibinfo {pages} {473} (\bibinfo {year} {2003})}\BibitemShut
  {NoStop}%
\bibitem [{\citenamefont {Shen}\ \emph {et~al.}(2005)\citenamefont {Shen},
  \citenamefont {Ronning}, \citenamefont {Lu}, \citenamefont {Baumberger},
  \citenamefont {Ingle}, \citenamefont {Lee}, \citenamefont {Meevasana},
  \citenamefont {Kohsaka}, \citenamefont {Azuma}, \citenamefont {Takano},
  \citenamefont {Takagi},\ and\ \citenamefont {Shen}}]{Shen2005}%
  \BibitemOpen
  \bibfield  {author} {\bibinfo {author} {\bibfnamefont {K.~M.}\ \bibnamefont
  {Shen}}, \bibinfo {author} {\bibfnamefont {F.}~\bibnamefont {Ronning}},
  \bibinfo {author} {\bibfnamefont {D.~H.}\ \bibnamefont {Lu}}, \bibinfo
  {author} {\bibfnamefont {F.}~\bibnamefont {Baumberger}}, \bibinfo {author}
  {\bibfnamefont {N.~J.~C.}\ \bibnamefont {Ingle}}, \bibinfo {author}
  {\bibfnamefont {W.~S.}\ \bibnamefont {Lee}}, \bibinfo {author} {\bibfnamefont
  {W.}~\bibnamefont {Meevasana}}, \bibinfo {author} {\bibfnamefont
  {Y.}~\bibnamefont {Kohsaka}}, \bibinfo {author} {\bibfnamefont
  {M.}~\bibnamefont {Azuma}}, \bibinfo {author} {\bibfnamefont
  {M.}~\bibnamefont {Takano}}, \bibinfo {author} {\bibfnamefont
  {H.}~\bibnamefont {Takagi}},\ and\ \bibinfo {author} {\bibfnamefont {Z.-X.}\
  \bibnamefont {Shen}},\ }\href {https://doi.org/10.1126/science.1103627}
  {\bibfield  {journal} {\bibinfo  {journal} {Science}\ }\textbf {\bibinfo
  {volume} {307}},\ \bibinfo {pages} {901} (\bibinfo {year}
  {2005})}\BibitemShut {NoStop}%
\bibitem [{\citenamefont {Chen}\ \emph {et~al.}(2019)\citenamefont {Chen},
  \citenamefont {Hashimoto}, \citenamefont {He}, \citenamefont {Song},
  \citenamefont {Xu}, \citenamefont {He}, \citenamefont {Devereaux},
  \citenamefont {Eisaki}, \citenamefont {Lu}, \citenamefont {Zaanen},\ and\
  \citenamefont {Shen}}]{Chen2019}%
  \BibitemOpen
  \bibfield  {author} {\bibinfo {author} {\bibfnamefont {S.-D.}\ \bibnamefont
  {Chen}}, \bibinfo {author} {\bibfnamefont {M.}~\bibnamefont {Hashimoto}},
  \bibinfo {author} {\bibfnamefont {Y.}~\bibnamefont {He}}, \bibinfo {author}
  {\bibfnamefont {D.}~\bibnamefont {Song}}, \bibinfo {author} {\bibfnamefont
  {K.-J.}\ \bibnamefont {Xu}}, \bibinfo {author} {\bibfnamefont {J.-F.}\
  \bibnamefont {He}}, \bibinfo {author} {\bibfnamefont {T.~P.}\ \bibnamefont
  {Devereaux}}, \bibinfo {author} {\bibfnamefont {H.}~\bibnamefont {Eisaki}},
  \bibinfo {author} {\bibfnamefont {D.-H.}\ \bibnamefont {Lu}}, \bibinfo
  {author} {\bibfnamefont {J.}~\bibnamefont {Zaanen}},\ and\ \bibinfo {author}
  {\bibfnamefont {Z.-X.}\ \bibnamefont {Shen}},\ }\href
  {https://doi.org/10.1126/science.aaw8850} {\bibfield  {journal} {\bibinfo
  {journal} {Science}\ }\textbf {\bibinfo {volume} {366}},\ \bibinfo {pages}
  {1099} (\bibinfo {year} {2019})}\BibitemShut {NoStop}%
\bibitem [{\citenamefont {Kunisada}\ \emph {et~al.}(2020)\citenamefont
  {Kunisada}, \citenamefont {Isono}, \citenamefont {Kohama}, \citenamefont
  {Sakai}, \citenamefont {Bareille}, \citenamefont {Sakuragi}, \citenamefont
  {Noguchi}, \citenamefont {Kurokawa}, \citenamefont {Kuroda}, \citenamefont
  {Ishida}, \citenamefont {Adachi}, \citenamefont {Sekine}, \citenamefont
  {Kim}, \citenamefont {Cacho}, \citenamefont {Shin}, \citenamefont {Tohyama},
  \citenamefont {Tokiwa},\ and\ \citenamefont {Kondo}}]{Kunisada2020}%
  \BibitemOpen
  \bibfield  {author} {\bibinfo {author} {\bibfnamefont {S.}~\bibnamefont
  {Kunisada}}, \bibinfo {author} {\bibfnamefont {S.}~\bibnamefont {Isono}},
  \bibinfo {author} {\bibfnamefont {Y.}~\bibnamefont {Kohama}}, \bibinfo
  {author} {\bibfnamefont {S.}~\bibnamefont {Sakai}}, \bibinfo {author}
  {\bibfnamefont {C.}~\bibnamefont {Bareille}}, \bibinfo {author}
  {\bibfnamefont {S.}~\bibnamefont {Sakuragi}}, \bibinfo {author}
  {\bibfnamefont {R.}~\bibnamefont {Noguchi}}, \bibinfo {author} {\bibfnamefont
  {K.}~\bibnamefont {Kurokawa}}, \bibinfo {author} {\bibfnamefont
  {K.}~\bibnamefont {Kuroda}}, \bibinfo {author} {\bibfnamefont
  {Y.}~\bibnamefont {Ishida}}, \bibinfo {author} {\bibfnamefont
  {S.}~\bibnamefont {Adachi}}, \bibinfo {author} {\bibfnamefont
  {R.}~\bibnamefont {Sekine}}, \bibinfo {author} {\bibfnamefont {T.~K.}\
  \bibnamefont {Kim}}, \bibinfo {author} {\bibfnamefont {C.}~\bibnamefont
  {Cacho}}, \bibinfo {author} {\bibfnamefont {S.}~\bibnamefont {Shin}},
  \bibinfo {author} {\bibfnamefont {T.}~\bibnamefont {Tohyama}}, \bibinfo
  {author} {\bibfnamefont {K.}~\bibnamefont {Tokiwa}},\ and\ \bibinfo {author}
  {\bibfnamefont {T.}~\bibnamefont {Kondo}},\ }\href
  {https://doi.org/10.1126/science.aay7311} {\bibfield  {journal} {\bibinfo
  {journal} {Science}\ }\textbf {\bibinfo {volume} {369}},\ \bibinfo {pages}
  {833} (\bibinfo {year} {2020})}\BibitemShut {NoStop}%
\bibitem [{\citenamefont {Scalapino}(1970)}]{Scalapino1970}%
  \BibitemOpen
  \bibfield  {author} {\bibinfo {author} {\bibfnamefont {D.~J.}\ \bibnamefont
  {Scalapino}},\ }\href {https://doi.org/10.1103/PhysRevLett.24.1052}
  {\bibfield  {journal} {\bibinfo  {journal} {Phys. Rev. Lett.}\ }\textbf
  {\bibinfo {volume} {24}},\ \bibinfo {pages} {1052} (\bibinfo {year}
  {1970})}\BibitemShut {NoStop}%
\bibitem [{\citenamefont {Anderson}\ \emph {et~al.}(1972)\citenamefont
  {Anderson}, \citenamefont {Carlson},\ and\ \citenamefont
  {Goldman}}]{Anderson1972}%
  \BibitemOpen
  \bibfield  {author} {\bibinfo {author} {\bibfnamefont {J.~T.}\ \bibnamefont
  {Anderson}}, \bibinfo {author} {\bibfnamefont {R.~V.}\ \bibnamefont
  {Carlson}},\ and\ \bibinfo {author} {\bibfnamefont {A.~M.}\ \bibnamefont
  {Goldman}},\ }\href {https://doi.org/10.1007/BF00655546} {\bibfield
  {journal} {\bibinfo  {journal} {Journal of Low Temperature Physics}\ }\textbf
  {\bibinfo {volume} {8}},\ \bibinfo {pages} {29} (\bibinfo {year}
  {1972})}\BibitemShut {NoStop}%
\bibitem [{\citenamefont {Bergeal}\ \emph {et~al.}(2008)\citenamefont
  {Bergeal}, \citenamefont {Lesueur}, \citenamefont {Aprili}, \citenamefont
  {Faini}, \citenamefont {Contour},\ and\ \citenamefont
  {Leridon}}]{Bergeal2008}%
  \BibitemOpen
  \bibfield  {author} {\bibinfo {author} {\bibfnamefont {N.}~\bibnamefont
  {Bergeal}}, \bibinfo {author} {\bibfnamefont {J.}~\bibnamefont {Lesueur}},
  \bibinfo {author} {\bibfnamefont {M.}~\bibnamefont {Aprili}}, \bibinfo
  {author} {\bibfnamefont {G.}~\bibnamefont {Faini}}, \bibinfo {author}
  {\bibfnamefont {J.~P.}\ \bibnamefont {Contour}},\ and\ \bibinfo {author}
  {\bibfnamefont {B.}~\bibnamefont {Leridon}},\ }\href
  {https://doi.org/10.1038/nphys1017} {\bibfield  {journal} {\bibinfo
  {journal} {Nature Phys}\ }\textbf {\bibinfo {volume} {4}},\ \bibinfo {pages}
  {608} (\bibinfo {year} {2008})}\BibitemShut {NoStop}%
\bibitem [{\citenamefont {Bastiaans}\ \emph {et~al.}(2021)\citenamefont
  {Bastiaans}, \citenamefont {Chatzopoulos}, \citenamefont {Ge}, \citenamefont
  {Cho}, \citenamefont {Tromp}, \citenamefont {{van Ruitenbeek}}, \citenamefont
  {Fischer}, \citenamefont {{de Visser}}, \citenamefont {Thoen}, \citenamefont
  {Driessen}, \citenamefont {Klapwijk},\ and\ \citenamefont
  {Allan}}]{Bastiaans2021}%
  \BibitemOpen
  \bibfield  {author} {\bibinfo {author} {\bibfnamefont {K.~M.}\ \bibnamefont
  {Bastiaans}}, \bibinfo {author} {\bibfnamefont {D.}~\bibnamefont
  {Chatzopoulos}}, \bibinfo {author} {\bibfnamefont {J.-F.}\ \bibnamefont
  {Ge}}, \bibinfo {author} {\bibfnamefont {D.}~\bibnamefont {Cho}}, \bibinfo
  {author} {\bibfnamefont {W.~O.}\ \bibnamefont {Tromp}}, \bibinfo {author}
  {\bibfnamefont {J.~M.}\ \bibnamefont {{van Ruitenbeek}}}, \bibinfo {author}
  {\bibfnamefont {M.~H.}\ \bibnamefont {Fischer}}, \bibinfo {author}
  {\bibfnamefont {P.~J.}\ \bibnamefont {{de Visser}}}, \bibinfo {author}
  {\bibfnamefont {D.~J.}\ \bibnamefont {Thoen}}, \bibinfo {author}
  {\bibfnamefont {E.~F.~C.}\ \bibnamefont {Driessen}}, \bibinfo {author}
  {\bibfnamefont {T.~M.}\ \bibnamefont {Klapwijk}},\ and\ \bibinfo {author}
  {\bibfnamefont {M.~P.}\ \bibnamefont {Allan}},\ }\href
  {https://doi.org/10.1126/science.abe3987} {\bibfield  {journal} {\bibinfo
  {journal} {Science}\ }\textbf {\bibinfo {volume} {374}},\ \bibinfo {pages}
  {608} (\bibinfo {year} {2021})}\BibitemShut {NoStop}%
\bibitem [{\citenamefont {Feshbach}(1958)}]{Feshbach1958}%
  \BibitemOpen
  \bibfield  {author} {\bibinfo {author} {\bibfnamefont {H.}~\bibnamefont
  {Feshbach}},\ }\href
  {https://doi.org/https://doi.org/10.1016/0003-4916(58)90007-1} {\bibfield
  {journal} {\bibinfo  {journal} {Annals of Physics}\ }\textbf {\bibinfo
  {volume} {5}},\ \bibinfo {pages} {357} (\bibinfo {year} {1958})}\BibitemShut
  {NoStop}%
\bibitem [{\citenamefont {Chen}\ \emph
  {et~al.}(2024{\natexlab{a}})\citenamefont {Chen}, \citenamefont {Wang},
  \citenamefont {Boyack}, \citenamefont {Yang},\ and\ \citenamefont
  {Levin}}]{Chen2024RMP}%
  \BibitemOpen
  \bibfield  {author} {\bibinfo {author} {\bibfnamefont {Q.}~\bibnamefont
  {Chen}}, \bibinfo {author} {\bibfnamefont {Z.}~\bibnamefont {Wang}}, \bibinfo
  {author} {\bibfnamefont {R.}~\bibnamefont {Boyack}}, \bibinfo {author}
  {\bibfnamefont {S.}~\bibnamefont {Yang}},\ and\ \bibinfo {author}
  {\bibfnamefont {K.}~\bibnamefont {Levin}},\ }\href
  {https://doi.org/10.1103/RevModPhys.96.025002} {\bibfield  {journal}
  {\bibinfo  {journal} {Rev. Mod. Phys.}\ }\textbf {\bibinfo {volume} {96}},\
  \bibinfo {pages} {025002} (\bibinfo {year} {2024}{\natexlab{a}})}\BibitemShut
  {NoStop}%
\bibitem [{\citenamefont {Homeier}\ \emph {et~al.}(2025)\citenamefont
  {Homeier}, \citenamefont {Lange}, \citenamefont {Demler}, \citenamefont
  {Bohrdt},\ and\ \citenamefont {Grusdt}}]{Homeier2025}%
  \BibitemOpen
  \bibfield  {author} {\bibinfo {author} {\bibfnamefont {L.}~\bibnamefont
  {Homeier}}, \bibinfo {author} {\bibfnamefont {H.}~\bibnamefont {Lange}},
  \bibinfo {author} {\bibfnamefont {E.}~\bibnamefont {Demler}}, \bibinfo
  {author} {\bibfnamefont {A.}~\bibnamefont {Bohrdt}},\ and\ \bibinfo {author}
  {\bibfnamefont {F.}~\bibnamefont {Grusdt}},\ }\href
  {https://doi.org/10.1038/s41467-024-55549-4} {\bibfield  {journal} {\bibinfo
  {journal} {Nature Communications}\ }\textbf {\bibinfo {volume} {16}},\
  \bibinfo {pages} {314} (\bibinfo {year} {2025})}\BibitemShut {NoStop}%
\bibitem [{\citenamefont {Bohrdt}\ \emph {et~al.}(2023)\citenamefont {Bohrdt},
  \citenamefont {Demler},\ and\ \citenamefont {Grusdt}}]{Bohrdt2023}%
  \BibitemOpen
  \bibfield  {author} {\bibinfo {author} {\bibfnamefont {A.}~\bibnamefont
  {Bohrdt}}, \bibinfo {author} {\bibfnamefont {E.}~\bibnamefont {Demler}},\
  and\ \bibinfo {author} {\bibfnamefont {F.}~\bibnamefont {Grusdt}},\ }\href
  {https://doi.org/10.1038/s41467-023-43453-2} {\bibfield  {journal} {\bibinfo
  {journal} {Nat Commun}\ }\textbf {\bibinfo {volume} {14}},\ \bibinfo {pages}
  {8017} (\bibinfo {year} {2023})}\BibitemShut {NoStop}%
\bibitem [{\citenamefont {Grusdt}\ \emph {et~al.}(2023)\citenamefont {Grusdt},
  \citenamefont {Demler},\ and\ \citenamefont {Bohrdt}}]{Grusdt2023}%
  \BibitemOpen
  \bibfield  {author} {\bibinfo {author} {\bibfnamefont {F.}~\bibnamefont
  {Grusdt}}, \bibinfo {author} {\bibfnamefont {E.}~\bibnamefont {Demler}},\
  and\ \bibinfo {author} {\bibfnamefont {A.}~\bibnamefont {Bohrdt}},\ }\href
  {https://doi.org/10.21468/SciPostPhys.14.5.090} {\bibfield  {journal}
  {\bibinfo  {journal} {SciPost Phys.}\ }\textbf {\bibinfo {volume} {14}},\
  \bibinfo {pages} {090} (\bibinfo {year} {2023})},\ \Eprint
  {https://arxiv.org/abs/2210.02321} {arXiv:2210.02321 [cond-mat]} \BibitemShut
  {NoStop}%
\bibitem [{\citenamefont {Homeier}\ \emph {et~al.}(2024)\citenamefont
  {Homeier}, \citenamefont {Bermes},\ and\ \citenamefont
  {Grusdt}}]{Homeier2024}%
  \BibitemOpen
  \bibfield  {author} {\bibinfo {author} {\bibfnamefont {L.}~\bibnamefont
  {Homeier}}, \bibinfo {author} {\bibfnamefont {P.}~\bibnamefont {Bermes}},\
  and\ \bibinfo {author} {\bibfnamefont {F.}~\bibnamefont {Grusdt}},\ }\href
  {https://doi.org/10.1103/PhysRevB.109.125135} {\bibfield  {journal} {\bibinfo
   {journal} {Phys. Rev. B}\ }\textbf {\bibinfo {volume} {109}},\ \bibinfo
  {pages} {125135} (\bibinfo {year} {2024})}\BibitemShut {NoStop}%
\bibitem [{\citenamefont {B{\'e}ran}\ \emph {et~al.}(1996)\citenamefont
  {B{\'e}ran}, \citenamefont {Poilblanc},\ and\ \citenamefont
  {Laughlin}}]{Beran1996}%
  \BibitemOpen
  \bibfield  {author} {\bibinfo {author} {\bibfnamefont {P.}~\bibnamefont
  {B{\'e}ran}}, \bibinfo {author} {\bibfnamefont {D.}~\bibnamefont
  {Poilblanc}},\ and\ \bibinfo {author} {\bibfnamefont {R.~B.}\ \bibnamefont
  {Laughlin}},\ }\href {https://doi.org/10.1016/0550-3213(96)00196-4}
  {\bibfield  {journal} {\bibinfo  {journal} {Nuclear Physics B}\ }\textbf
  {\bibinfo {volume} {473}},\ \bibinfo {pages} {707} (\bibinfo {year}
  {1996})}\BibitemShut {NoStop}%
\bibitem [{\citenamefont {Grusdt}\ \emph {et~al.}(2018)\citenamefont {Grusdt},
  \citenamefont {{Kan{\'a}sz-Nagy}}, \citenamefont {Bohrdt}, \citenamefont
  {Chiu}, \citenamefont {Ji}, \citenamefont {Greiner}, \citenamefont {Greif},\
  and\ \citenamefont {Demler}}]{Grusdt2018}%
  \BibitemOpen
  \bibfield  {author} {\bibinfo {author} {\bibfnamefont {F.}~\bibnamefont
  {Grusdt}}, \bibinfo {author} {\bibfnamefont {M.}~\bibnamefont
  {{Kan{\'a}sz-Nagy}}}, \bibinfo {author} {\bibfnamefont {A.}~\bibnamefont
  {Bohrdt}}, \bibinfo {author} {\bibfnamefont {C.~S.}\ \bibnamefont {Chiu}},
  \bibinfo {author} {\bibfnamefont {G.}~\bibnamefont {Ji}}, \bibinfo {author}
  {\bibfnamefont {M.}~\bibnamefont {Greiner}}, \bibinfo {author} {\bibfnamefont
  {D.}~\bibnamefont {Greif}},\ and\ \bibinfo {author} {\bibfnamefont
  {E.}~\bibnamefont {Demler}},\ }\href
  {https://doi.org/10.1103/PhysRevX.8.011046} {\bibfield  {journal} {\bibinfo
  {journal} {Phys. Rev. X}\ }\textbf {\bibinfo {volume} {8}},\ \bibinfo {pages}
  {011046} (\bibinfo {year} {2018})}\BibitemShut {NoStop}%
\bibitem [{\citenamefont {Bohrdt}\ \emph
  {et~al.}(2021{\natexlab{a}})\citenamefont {Bohrdt}, \citenamefont {Demler},\
  and\ \citenamefont {Grusdt}}]{Bohrdt2021}%
  \BibitemOpen
  \bibfield  {author} {\bibinfo {author} {\bibfnamefont {A.}~\bibnamefont
  {Bohrdt}}, \bibinfo {author} {\bibfnamefont {E.}~\bibnamefont {Demler}},\
  and\ \bibinfo {author} {\bibfnamefont {F.}~\bibnamefont {Grusdt}},\ }\href
  {https://doi.org/10.1103/PhysRevLett.127.197004} {\bibfield  {journal}
  {\bibinfo  {journal} {Phys. Rev. Lett.}\ }\textbf {\bibinfo {volume} {127}},\
  \bibinfo {pages} {197004} (\bibinfo {year} {2021}{\natexlab{a}})}\BibitemShut
  {NoStop}%
\bibitem [{\citenamefont {Sous}\ \emph {et~al.}(2023)\citenamefont {Sous},
  \citenamefont {He},\ and\ \citenamefont {Kivelson}}]{Sous2023}%
  \BibitemOpen
  \bibfield  {author} {\bibinfo {author} {\bibfnamefont {J.}~\bibnamefont
  {Sous}}, \bibinfo {author} {\bibfnamefont {Y.}~\bibnamefont {He}},\ and\
  \bibinfo {author} {\bibfnamefont {S.~A.}\ \bibnamefont {Kivelson}},\ }\href
  {https://doi.org/10.1038/s41535-023-00550-1} {\bibfield  {journal} {\bibinfo
  {journal} {npj Quantum Mater.}\ }\textbf {\bibinfo {volume} {8}},\ \bibinfo
  {pages} {25} (\bibinfo {year} {2023})},\ \Eprint
  {https://arxiv.org/abs/2210.13478} {arXiv:2210.13478 [cond-mat]} \BibitemShut
  {NoStop}%
\bibitem [{\citenamefont {Chen}\ \emph
  {et~al.}(2024{\natexlab{b}})\citenamefont {Chen}, \citenamefont {Wang},
  \citenamefont {Boyack},\ and\ \citenamefont {Levin}}]{Chen2024test}%
  \BibitemOpen
  \bibfield  {author} {\bibinfo {author} {\bibfnamefont {Q.}~\bibnamefont
  {Chen}}, \bibinfo {author} {\bibfnamefont {Z.}~\bibnamefont {Wang}}, \bibinfo
  {author} {\bibfnamefont {R.}~\bibnamefont {Boyack}},\ and\ \bibinfo {author}
  {\bibfnamefont {K.}~\bibnamefont {Levin}},\ }\href
  {https://doi.org/10.1038/s41535-024-00640-8} {\bibfield  {journal} {\bibinfo
  {journal} {npj Quantum Materials}\ }\textbf {\bibinfo {volume} {9}},\
  \bibinfo {pages} {27} (\bibinfo {year} {2024}{\natexlab{b}})}\BibitemShut
  {NoStop}%
\bibitem [{\citenamefont {Paeckel}(2024)}]{Paeckel2024}%
  \BibitemOpen
  \bibfield  {author} {\bibinfo {author} {\bibfnamefont {S.}~\bibnamefont
  {Paeckel}},\ }\href {https://doi.org/10.48550/arXiv.2411.09680} {\bibinfo
  {title} {Spectral decomposition and high-accuracy {{Greens}} functions:
  {{Overcoming}} the {{Nyquist-Shannon}} limit via complex-time {{Krylov}}
  expansion}} (\bibinfo {year} {2024}),\ \Eprint
  {https://arxiv.org/abs/2411.09680} {arXiv:2411.09680 [cond-mat]} \BibitemShut
  {NoStop}%
\bibitem [{\citenamefont {Paeckel}\ \emph {et~al.}(2019)\citenamefont
  {Paeckel}, \citenamefont {K{\"o}hler}, \citenamefont {Swoboda}, \citenamefont
  {Manmana}, \citenamefont {Schollw{\"o}ck},\ and\ \citenamefont
  {Hubig}}]{PAECKEL2019}%
  \BibitemOpen
  \bibfield  {author} {\bibinfo {author} {\bibfnamefont {S.}~\bibnamefont
  {Paeckel}}, \bibinfo {author} {\bibfnamefont {T.}~\bibnamefont {K{\"o}hler}},
  \bibinfo {author} {\bibfnamefont {A.}~\bibnamefont {Swoboda}}, \bibinfo
  {author} {\bibfnamefont {S.~R.}\ \bibnamefont {Manmana}}, \bibinfo {author}
  {\bibfnamefont {U.}~\bibnamefont {Schollw{\"o}ck}},\ and\ \bibinfo {author}
  {\bibfnamefont {C.}~\bibnamefont {Hubig}},\ }\href
  {https://doi.org/https://doi.org/10.1016/j.aop.2019.167998} {\bibfield
  {journal} {\bibinfo  {journal} {Annals of Physics}\ }\textbf {\bibinfo
  {volume} {411}},\ \bibinfo {pages} {167998} (\bibinfo {year}
  {2019})}\BibitemShut {NoStop}%
\bibitem [{\citenamefont {Zaletel}\ \emph {et~al.}(2015)\citenamefont
  {Zaletel}, \citenamefont {Mong}, \citenamefont {Karrasch}, \citenamefont
  {Moore},\ and\ \citenamefont {Pollmann}}]{Zaletel2015}%
  \BibitemOpen
  \bibfield  {author} {\bibinfo {author} {\bibfnamefont {M.~P.}\ \bibnamefont
  {Zaletel}}, \bibinfo {author} {\bibfnamefont {R.~S.~K.}\ \bibnamefont
  {Mong}}, \bibinfo {author} {\bibfnamefont {C.}~\bibnamefont {Karrasch}},
  \bibinfo {author} {\bibfnamefont {J.~E.}\ \bibnamefont {Moore}},\ and\
  \bibinfo {author} {\bibfnamefont {F.}~\bibnamefont {Pollmann}},\ }\href
  {https://doi.org/10.1103/PhysRevB.91.165112} {\bibfield  {journal} {\bibinfo
  {journal} {Phys. Rev. B}\ }\textbf {\bibinfo {volume} {91}},\ \bibinfo
  {pages} {165112} (\bibinfo {year} {2015})}\BibitemShut {NoStop}%
\bibitem [{\citenamefont {Shraiman}\ and\ \citenamefont
  {Siggia}(1988{\natexlab{a}})}]{Shraiman1988a}%
  \BibitemOpen
  \bibfield  {author} {\bibinfo {author} {\bibfnamefont {B.~I.}\ \bibnamefont
  {Shraiman}}\ and\ \bibinfo {author} {\bibfnamefont {E.~D.}\ \bibnamefont
  {Siggia}},\ }\href {https://doi.org/10.1103/PhysRevLett.61.467} {\bibfield
  {journal} {\bibinfo  {journal} {Phys. Rev. Lett.}\ }\textbf {\bibinfo
  {volume} {61}},\ \bibinfo {pages} {467} (\bibinfo {year}
  {1988}{\natexlab{a}})}\BibitemShut {NoStop}%
\bibitem [{\citenamefont {Vidmar}\ and\ \citenamefont
  {Bonca}(2013)}]{Vidmar2013}%
  \BibitemOpen
  \bibfield  {author} {\bibinfo {author} {\bibfnamefont {L.}~\bibnamefont
  {Vidmar}}\ and\ \bibinfo {author} {\bibfnamefont {J.}~\bibnamefont {Bonca}},\
  }\href {https://doi.org/10.1007/s10948-013-2151-2} {\bibfield  {journal}
  {\bibinfo  {journal} {J Supercond Nov Magn}\ }\textbf {\bibinfo {volume}
  {26}},\ \bibinfo {pages} {2641} (\bibinfo {year} {2013})},\ \Eprint
  {https://arxiv.org/abs/1311.5283} {arXiv:1311.5283 [cond-mat]} \BibitemShut
  {NoStop}%
\bibitem [{\citenamefont {Grusdt}\ \emph {et~al.}(2019)\citenamefont {Grusdt},
  \citenamefont {Bohrdt},\ and\ \citenamefont {Demler}}]{Grusdt2019PRB}%
  \BibitemOpen
  \bibfield  {author} {\bibinfo {author} {\bibfnamefont {F.}~\bibnamefont
  {Grusdt}}, \bibinfo {author} {\bibfnamefont {A.}~\bibnamefont {Bohrdt}},\
  and\ \bibinfo {author} {\bibfnamefont {E.}~\bibnamefont {Demler}},\ }\href
  {https://doi.org/10.1103/PhysRevB.99.224422} {\bibfield  {journal} {\bibinfo
  {journal} {Phys. Rev. B}\ }\textbf {\bibinfo {volume} {99}},\ \bibinfo
  {pages} {224422} (\bibinfo {year} {2019})}\BibitemShut {NoStop}%
\bibitem [{\citenamefont {Bohrdt}\ \emph {et~al.}(2020)\citenamefont {Bohrdt},
  \citenamefont {Demler}, \citenamefont {Pollmann}, \citenamefont {Knap},\ and\
  \citenamefont {Grusdt}}]{Bohrdt2020}%
  \BibitemOpen
  \bibfield  {author} {\bibinfo {author} {\bibfnamefont {A.}~\bibnamefont
  {Bohrdt}}, \bibinfo {author} {\bibfnamefont {E.}~\bibnamefont {Demler}},
  \bibinfo {author} {\bibfnamefont {F.}~\bibnamefont {Pollmann}}, \bibinfo
  {author} {\bibfnamefont {M.}~\bibnamefont {Knap}},\ and\ \bibinfo {author}
  {\bibfnamefont {F.}~\bibnamefont {Grusdt}},\ }\href
  {https://doi.org/10.1103/PhysRevB.102.035139} {\bibfield  {journal} {\bibinfo
   {journal} {Phys. Rev. B}\ }\textbf {\bibinfo {volume} {102}},\ \bibinfo
  {pages} {035139} (\bibinfo {year} {2020})}\BibitemShut {NoStop}%
\bibitem [{\citenamefont {Shraiman}\ and\ \citenamefont
  {Siggia}(1988{\natexlab{b}})}]{Shraiman1988b}%
  \BibitemOpen
  \bibfield  {author} {\bibinfo {author} {\bibfnamefont {B.~I.}\ \bibnamefont
  {Shraiman}}\ and\ \bibinfo {author} {\bibfnamefont {E.~D.}\ \bibnamefont
  {Siggia}},\ }\href {https://doi.org/10.1103/PhysRevLett.60.740} {\bibfield
  {journal} {\bibinfo  {journal} {Phys. Rev. Lett.}\ }\textbf {\bibinfo
  {volume} {60}},\ \bibinfo {pages} {740} (\bibinfo {year}
  {1988}{\natexlab{b}})}\BibitemShut {NoStop}%
\bibitem [{\citenamefont {Dalla~Piazza}\ \emph {et~al.}(2015)\citenamefont
  {Dalla~Piazza}, \citenamefont {Mourigal}, \citenamefont {Christensen},
  \citenamefont {Nilsen}, \citenamefont {Tregenna-Piggott}, \citenamefont
  {Perring}, \citenamefont {Enderle}, \citenamefont {McMorrow}, \citenamefont
  {Ivanov},\ and\ \citenamefont {Ronnow}}]{Piazza2015}%
  \BibitemOpen
  \bibfield  {author} {\bibinfo {author} {\bibfnamefont {B.}~\bibnamefont
  {Dalla~Piazza}}, \bibinfo {author} {\bibfnamefont {M.}~\bibnamefont
  {Mourigal}}, \bibinfo {author} {\bibfnamefont {N.~B.}\ \bibnamefont
  {Christensen}}, \bibinfo {author} {\bibfnamefont {G.~J.}\ \bibnamefont
  {Nilsen}}, \bibinfo {author} {\bibfnamefont {P.}~\bibnamefont
  {Tregenna-Piggott}}, \bibinfo {author} {\bibfnamefont {T.~G.}\ \bibnamefont
  {Perring}}, \bibinfo {author} {\bibfnamefont {M.}~\bibnamefont {Enderle}},
  \bibinfo {author} {\bibfnamefont {D.~F.}\ \bibnamefont {McMorrow}}, \bibinfo
  {author} {\bibfnamefont {D.~A.}\ \bibnamefont {Ivanov}},\ and\ \bibinfo
  {author} {\bibfnamefont {H.~M.}\ \bibnamefont {Ronnow}},\ }\href
  {https://doi.org/10.1038/nphys3172} {\bibfield  {journal} {\bibinfo
  {journal} {Nat Phys}\ }\textbf {\bibinfo {volume} {11}},\ \bibinfo {pages}
  {62} (\bibinfo {year} {2015})}\BibitemShut {NoStop}%
\bibitem [{\citenamefont {Manousakis}(2007)}]{Manousakis2007}%
  \BibitemOpen
  \bibfield  {author} {\bibinfo {author} {\bibfnamefont {E.}~\bibnamefont
  {Manousakis}},\ }\href {https://doi.org/10.1103/PhysRevB.75.035106}
  {\bibfield  {journal} {\bibinfo  {journal} {Phys. Rev. B}\ }\textbf {\bibinfo
  {volume} {75}},\ \bibinfo {pages} {035106} (\bibinfo {year}
  {2007})}\BibitemShut {NoStop}%
\bibitem [{\citenamefont {Simons}\ and\ \citenamefont
  {Gunn}(1990)}]{Simons1990}%
  \BibitemOpen
  \bibfield  {author} {\bibinfo {author} {\bibfnamefont {B.~D.}\ \bibnamefont
  {Simons}}\ and\ \bibinfo {author} {\bibfnamefont {J.~M.~F.}\ \bibnamefont
  {Gunn}},\ }\href {https://doi.org/10.1103/PhysRevB.41.7019} {\bibfield
  {journal} {\bibinfo  {journal} {Phys. Rev. B}\ }\textbf {\bibinfo {volume}
  {41}},\ \bibinfo {pages} {7019} (\bibinfo {year} {1990})}\BibitemShut
  {NoStop}%
\bibitem [{\citenamefont {Bermes}\ \emph {et~al.}(2024)\citenamefont {Bermes},
  \citenamefont {Bohrdt},\ and\ \citenamefont {Grusdt}}]{Bermes2024}%
  \BibitemOpen
  \bibfield  {author} {\bibinfo {author} {\bibfnamefont {P.}~\bibnamefont
  {Bermes}}, \bibinfo {author} {\bibfnamefont {A.}~\bibnamefont {Bohrdt}},\
  and\ \bibinfo {author} {\bibfnamefont {F.}~\bibnamefont {Grusdt}},\ }\href
  {https://doi.org/10.1103/PhysRevB.109.205104} {\bibfield  {journal} {\bibinfo
   {journal} {Phys. Rev. B}\ }\textbf {\bibinfo {volume} {109}},\ \bibinfo
  {pages} {205104} (\bibinfo {year} {2024})}\BibitemShut {NoStop}%
\bibitem [{\citenamefont {Berakdar}(1998)}]{Berakdar1998}%
  \BibitemOpen
  \bibfield  {author} {\bibinfo {author} {\bibfnamefont {J.}~\bibnamefont
  {Berakdar}},\ }\href {https://doi.org/10.1103/PhysRevB.58.9808} {\bibfield
  {journal} {\bibinfo  {journal} {Phys. Rev. B}\ }\textbf {\bibinfo {volume}
  {58}},\ \bibinfo {pages} {9808} (\bibinfo {year} {1998})}\BibitemShut
  {NoStop}%
\bibitem [{\citenamefont {Su}\ and\ \citenamefont {Zhang}(2020)}]{Su2020}%
  \BibitemOpen
  \bibfield  {author} {\bibinfo {author} {\bibfnamefont {Y.}~\bibnamefont
  {Su}}\ and\ \bibinfo {author} {\bibfnamefont {C.}~\bibnamefont {Zhang}},\
  }\href {https://doi.org/10.1103/PhysRevB.101.205110} {\bibfield  {journal}
  {\bibinfo  {journal} {Phys. Rev. B}\ }\textbf {\bibinfo {volume} {101}},\
  \bibinfo {pages} {205110} (\bibinfo {year} {2020})}\BibitemShut {NoStop}%
\bibitem [{\citenamefont {Mahmood}\ \emph {et~al.}(2022)\citenamefont
  {Mahmood}, \citenamefont {Devereaux}, \citenamefont {Abbamonte},\ and\
  \citenamefont {Morr}}]{Mahmood2022}%
  \BibitemOpen
  \bibfield  {author} {\bibinfo {author} {\bibfnamefont {F.}~\bibnamefont
  {Mahmood}}, \bibinfo {author} {\bibfnamefont {T.}~\bibnamefont {Devereaux}},
  \bibinfo {author} {\bibfnamefont {P.}~\bibnamefont {Abbamonte}},\ and\
  \bibinfo {author} {\bibfnamefont {D.~K.}\ \bibnamefont {Morr}},\ }\href
  {https://doi.org/10.1103/PhysRevB.105.064515} {\bibfield  {journal} {\bibinfo
   {journal} {Phys. Rev. B}\ }\textbf {\bibinfo {volume} {105}},\ \bibinfo
  {pages} {064515} (\bibinfo {year} {2022})}\BibitemShut {NoStop}%
\bibitem [{\citenamefont {Kemper}\ \emph {et~al.}(2025)\citenamefont {Kemper},
  \citenamefont {Goto}, \citenamefont {Labib}, \citenamefont {Gauthier},
  \citenamefont {Neto},\ and\ \citenamefont {Boschini}}]{Kemper2025}%
  \BibitemOpen
  \bibfield  {author} {\bibinfo {author} {\bibfnamefont {A.~F.}\ \bibnamefont
  {Kemper}}, \bibinfo {author} {\bibfnamefont {F.}~\bibnamefont {Goto}},
  \bibinfo {author} {\bibfnamefont {H.~A.}\ \bibnamefont {Labib}}, \bibinfo
  {author} {\bibfnamefont {N.}~\bibnamefont {Gauthier}}, \bibinfo {author}
  {\bibfnamefont {E.~H. d.~S.}\ \bibnamefont {Neto}},\ and\ \bibinfo {author}
  {\bibfnamefont {F.}~\bibnamefont {Boschini}},\ }\href
  {https://doi.org/10.1103/4fqm-z742} {\bibfield  {journal} {\bibinfo
  {journal} {Phys. Rev. B}\ }\textbf {\bibinfo {volume} {112}},\ \bibinfo
  {pages} {035168} (\bibinfo {year} {2025})}\BibitemShut {NoStop}%
\bibitem [{\citenamefont {Tarruell}\ and\ \citenamefont
  {Sanchez-Palencia}()}]{Tarruell2018}%
  \BibitemOpen
  \bibfield  {author} {\bibinfo {author} {\bibfnamefont {L.}~\bibnamefont
  {Tarruell}}\ and\ \bibinfo {author} {\bibfnamefont {L.}~\bibnamefont
  {Sanchez-Palencia}},\ }\href@noop {} {\ }\Eprint
  {https://arxiv.org/abs/1809.00571} {1809.00571} \BibitemShut {NoStop}%
\bibitem [{\citenamefont {Gross}\ and\ \citenamefont
  {Bloch}(2017)}]{Gross2017}%
  \BibitemOpen
  \bibfield  {author} {\bibinfo {author} {\bibfnamefont {C.}~\bibnamefont
  {Gross}}\ and\ \bibinfo {author} {\bibfnamefont {I.}~\bibnamefont {Bloch}},\
  }\href {https://doi.org/10.1126/science.aal3837} {\bibfield  {journal}
  {\bibinfo  {journal} {Science}\ }\textbf {\bibinfo {volume} {357}},\ \bibinfo
  {pages} {995} (\bibinfo {year} {2017})}\BibitemShut {NoStop}%
\bibitem [{\citenamefont {Bohrdt}\ \emph
  {et~al.}(2021{\natexlab{b}})\citenamefont {Bohrdt}, \citenamefont {Homeier},
  \citenamefont {Reinmoser}, \citenamefont {Demler},\ and\ \citenamefont
  {Grusdt}}]{Bohrdt2021coldatoms}%
  \BibitemOpen
  \bibfield  {author} {\bibinfo {author} {\bibfnamefont {A.}~\bibnamefont
  {Bohrdt}}, \bibinfo {author} {\bibfnamefont {L.}~\bibnamefont {Homeier}},
  \bibinfo {author} {\bibfnamefont {C.}~\bibnamefont {Reinmoser}}, \bibinfo
  {author} {\bibfnamefont {E.}~\bibnamefont {Demler}},\ and\ \bibinfo {author}
  {\bibfnamefont {F.}~\bibnamefont {Grusdt}},\ }\href
  {https://doi.org/10.1016/j.aop.2021.168651} {\bibfield  {journal} {\bibinfo
  {journal} {Annals of Physics}\ }\bibinfo {series} {Special Issue on {{Philip
  W}}. {{Anderson}}},\ \textbf {\bibinfo {volume} {435}},\ \bibinfo {pages}
  {168651} (\bibinfo {year} {2021}{\natexlab{b}})}\BibitemShut {NoStop}%
\bibitem [{\citenamefont {Ho}\ \emph {et~al.}(2009)\citenamefont {Ho},
  \citenamefont {Cazalilla},\ and\ \citenamefont {Giamarchi}}]{Ho2009}%
  \BibitemOpen
  \bibfield  {author} {\bibinfo {author} {\bibfnamefont {A.~F.}\ \bibnamefont
  {Ho}}, \bibinfo {author} {\bibfnamefont {M.~A.}\ \bibnamefont {Cazalilla}},\
  and\ \bibinfo {author} {\bibfnamefont {T.}~\bibnamefont {Giamarchi}},\ }\href
  {https://doi.org/10.1103/PhysRevA.79.033620} {\bibfield  {journal} {\bibinfo
  {journal} {Phys. Rev. A}\ }\textbf {\bibinfo {volume} {79}},\ \bibinfo
  {pages} {033620} (\bibinfo {year} {2009})}\BibitemShut {NoStop}%
\bibitem [{\citenamefont {Blatz}\ \emph {et~al.}(2025)\citenamefont {Blatz},
  \citenamefont {Schollw\"ock}, \citenamefont {Grusdt},\ and\ \citenamefont
  {Bohrdt}}]{Blatz2025}%
  \BibitemOpen
  \bibfield  {author} {\bibinfo {author} {\bibfnamefont {T.}~\bibnamefont
  {Blatz}}, \bibinfo {author} {\bibfnamefont {U.}~\bibnamefont {Schollw\"ock}},
  \bibinfo {author} {\bibfnamefont {F.}~\bibnamefont {Grusdt}},\ and\ \bibinfo
  {author} {\bibfnamefont {A.}~\bibnamefont {Bohrdt}},\ }\href
  {https://doi.org/10.1103/dpfl-12st} {\bibfield  {journal} {\bibinfo
  {journal} {Phys. Rev. X}\ }\textbf {\bibinfo {volume} {15}},\ \bibinfo
  {pages} {031074} (\bibinfo {year} {2025})}\BibitemShut {NoStop}%
\bibitem [{\citenamefont {{Simons Collaboration on the Many-Electron Problem}}\
  \emph {et~al.}(2020)\citenamefont {{Simons Collaboration on the Many-Electron
  Problem}}, \citenamefont {Qin}, \citenamefont {Chung}, \citenamefont {Shi},
  \citenamefont {Vitali}, \citenamefont {Hubig}, \citenamefont
  {Schollw{\"o}ck}, \citenamefont {White},\ and\ \citenamefont
  {Zhang}}]{Qin2020}%
  \BibitemOpen
  \bibfield  {author} {\bibinfo {author} {\bibnamefont {{Simons Collaboration
  on the Many-Electron Problem}}}, \bibinfo {author} {\bibfnamefont
  {M.}~\bibnamefont {Qin}}, \bibinfo {author} {\bibfnamefont {C.-M.}\
  \bibnamefont {Chung}}, \bibinfo {author} {\bibfnamefont {H.}~\bibnamefont
  {Shi}}, \bibinfo {author} {\bibfnamefont {E.}~\bibnamefont {Vitali}},
  \bibinfo {author} {\bibfnamefont {C.}~\bibnamefont {Hubig}}, \bibinfo
  {author} {\bibfnamefont {U.}~\bibnamefont {Schollw{\"o}ck}}, \bibinfo
  {author} {\bibfnamefont {S.~R.}\ \bibnamefont {White}},\ and\ \bibinfo
  {author} {\bibfnamefont {S.}~\bibnamefont {Zhang}},\ }\href
  {https://doi.org/10.1103/PhysRevX.10.031016} {\bibfield  {journal} {\bibinfo
  {journal} {Phys. Rev. X}\ }\textbf {\bibinfo {volume} {10}},\ \bibinfo
  {pages} {031016} (\bibinfo {year} {2020})}\BibitemShut {NoStop}%
\bibitem [{\citenamefont {Xu}\ \emph {et~al.}(2025)\citenamefont {Xu},
  \citenamefont {Kendrick}, \citenamefont {Kale}, \citenamefont {Gang},
  \citenamefont {Feng}, \citenamefont {Zhang}, \citenamefont {Young},
  \citenamefont {Lebrat},\ and\ \citenamefont {Greiner}}]{Xu2025a}%
  \BibitemOpen
  \bibfield  {author} {\bibinfo {author} {\bibfnamefont {M.}~\bibnamefont
  {Xu}}, \bibinfo {author} {\bibfnamefont {L.~H.}\ \bibnamefont {Kendrick}},
  \bibinfo {author} {\bibfnamefont {A.}~\bibnamefont {Kale}}, \bibinfo {author}
  {\bibfnamefont {Y.}~\bibnamefont {Gang}}, \bibinfo {author} {\bibfnamefont
  {C.}~\bibnamefont {Feng}}, \bibinfo {author} {\bibfnamefont {S.}~\bibnamefont
  {Zhang}}, \bibinfo {author} {\bibfnamefont {A.~W.}\ \bibnamefont {Young}},
  \bibinfo {author} {\bibfnamefont {M.}~\bibnamefont {Lebrat}},\ and\ \bibinfo
  {author} {\bibfnamefont {M.}~\bibnamefont {Greiner}},\ }\href
  {https://doi.org/10.1038/s41586-025-09112-w} {\bibfield  {journal} {\bibinfo
  {journal} {Nature}\ }\textbf {\bibinfo {volume} {642}},\ \bibinfo {pages}
  {909} (\bibinfo {year} {2025})}\BibitemShut {NoStop}%
\bibitem [{\citenamefont {Roth}\ \emph {et~al.}(2025)\citenamefont {Roth},
  \citenamefont {Chen}, \citenamefont {Sengupta},\ and\ \citenamefont
  {Georges}}]{roth2025arXiv}%
  \BibitemOpen
  \bibfield  {author} {\bibinfo {author} {\bibfnamefont {C.}~\bibnamefont
  {Roth}}, \bibinfo {author} {\bibfnamefont {A.}~\bibnamefont {Chen}}, \bibinfo
  {author} {\bibfnamefont {A.}~\bibnamefont {Sengupta}},\ and\ \bibinfo
  {author} {\bibfnamefont {A.}~\bibnamefont {Georges}},\ }\href
  {https://arxiv.org/abs/2511.07566} {\bibinfo {title} {Superconductivity in
  the two-dimensional hubbard model revealed by neural quantum states}}
  (\bibinfo {year} {2025}),\ \Eprint {https://arxiv.org/abs/2511.07566}
  {arXiv:2511.07566 [cond-mat.supr-con]} \BibitemShut {NoStop}%
\bibitem [{\citenamefont {Lange}\ \emph {et~al.}(2024)\citenamefont {Lange},
  \citenamefont {Van~de Walle}, \citenamefont {Abedinnia},\ and\ \citenamefont
  {Bohrdt}}]{Lange2024}%
  \BibitemOpen
  \bibfield  {author} {\bibinfo {author} {\bibfnamefont {H.}~\bibnamefont
  {Lange}}, \bibinfo {author} {\bibfnamefont {A.}~\bibnamefont {Van~de Walle}},
  \bibinfo {author} {\bibfnamefont {A.}~\bibnamefont {Abedinnia}},\ and\
  \bibinfo {author} {\bibfnamefont {A.}~\bibnamefont {Bohrdt}},\ }\href
  {https://doi.org/10.1088/2058-9565/ad7168} {\bibfield  {journal} {\bibinfo
  {journal} {Quantum Science and Technology}\ }\textbf {\bibinfo {volume}
  {9}},\ \bibinfo {pages} {040501} (\bibinfo {year} {2024})}\BibitemShut
  {NoStop}%
\bibitem [{\citenamefont {Van~de Walle}\ \emph {et~al.}(2025)\citenamefont
  {Van~de Walle}, \citenamefont {Schmitt},\ and\ \citenamefont
  {Bohrdt}}]{VandeWalle2025}%
  \BibitemOpen
  \bibfield  {author} {\bibinfo {author} {\bibfnamefont {A.}~\bibnamefont
  {Van~de Walle}}, \bibinfo {author} {\bibfnamefont {M.}~\bibnamefont
  {Schmitt}},\ and\ \bibinfo {author} {\bibfnamefont {A.}~\bibnamefont
  {Bohrdt}},\ }\href {https://doi.org/10.1088/2632-2153/ae0f39} {\bibfield
  {journal} {\bibinfo  {journal} {Machine Learning: Science and Technology}\
  }\textbf {\bibinfo {volume} {6}},\ \bibinfo {pages} {045011} (\bibinfo {year}
  {2025})}\BibitemShut {NoStop}%
\bibitem [{\citenamefont {Micnas}\ \emph {et~al.}(1990)\citenamefont {Micnas},
  \citenamefont {Ranninger},\ and\ \citenamefont {Robaszkiewicz}}]{Micnas1990}%
  \BibitemOpen
  \bibfield  {author} {\bibinfo {author} {\bibfnamefont {R.}~\bibnamefont
  {Micnas}}, \bibinfo {author} {\bibfnamefont {J.}~\bibnamefont {Ranninger}},\
  and\ \bibinfo {author} {\bibfnamefont {S.}~\bibnamefont {Robaszkiewicz}},\
  }\href {https://doi.org/10.1103/RevModPhys.62.113} {\bibfield  {journal}
  {\bibinfo  {journal} {Rev. Mod. Phys.}\ }\textbf {\bibinfo {volume} {62}},\
  \bibinfo {pages} {113} (\bibinfo {year} {1990})}\BibitemShut {NoStop}%
\bibitem [{\citenamefont {Moreo}\ and\ \citenamefont
  {Scalapino}(2007)}]{Moreo2007}%
  \BibitemOpen
  \bibfield  {author} {\bibinfo {author} {\bibfnamefont {A.}~\bibnamefont
  {Moreo}}\ and\ \bibinfo {author} {\bibfnamefont {D.~J.}\ \bibnamefont
  {Scalapino}},\ }\href {https://doi.org/10.1103/PhysRevLett.98.216402}
  {\bibfield  {journal} {\bibinfo  {journal} {Phys. Rev. Lett.}\ }\textbf
  {\bibinfo {volume} {98}},\ \bibinfo {pages} {216402} (\bibinfo {year}
  {2007})}\BibitemShut {NoStop}%
\bibitem [{\citenamefont {Mitra}\ \emph {et~al.}(2018)\citenamefont {Mitra},
  \citenamefont {Brown}, \citenamefont {{Guardado-Sanchez}}, \citenamefont
  {Kondov}, \citenamefont {Devakul}, \citenamefont {Huse}, \citenamefont
  {Schau{\ss}},\ and\ \citenamefont {Bakr}}]{Mitra2018}%
  \BibitemOpen
  \bibfield  {author} {\bibinfo {author} {\bibfnamefont {D.}~\bibnamefont
  {Mitra}}, \bibinfo {author} {\bibfnamefont {P.~T.}\ \bibnamefont {Brown}},
  \bibinfo {author} {\bibfnamefont {E.}~\bibnamefont {{Guardado-Sanchez}}},
  \bibinfo {author} {\bibfnamefont {S.~S.}\ \bibnamefont {Kondov}}, \bibinfo
  {author} {\bibfnamefont {T.}~\bibnamefont {Devakul}}, \bibinfo {author}
  {\bibfnamefont {D.~A.}\ \bibnamefont {Huse}}, \bibinfo {author}
  {\bibfnamefont {P.}~\bibnamefont {Schau{\ss}}},\ and\ \bibinfo {author}
  {\bibfnamefont {W.~S.}\ \bibnamefont {Bakr}},\ }\href
  {https://doi.org/10.1038/nphys4297} {\bibfield  {journal} {\bibinfo
  {journal} {Nat. Phys.}\ }\textbf {\bibinfo {volume} {14}},\ \bibinfo {pages}
  {173} (\bibinfo {year} {2018})}\BibitemShut {NoStop}%
\bibitem [{\citenamefont {Gall}\ \emph {et~al.}(2020)\citenamefont {Gall},
  \citenamefont {Chan}, \citenamefont {Wurz},\ and\ \citenamefont
  {K{\"o}hl}}]{Gall2020}%
  \BibitemOpen
  \bibfield  {author} {\bibinfo {author} {\bibfnamefont {M.}~\bibnamefont
  {Gall}}, \bibinfo {author} {\bibfnamefont {C.~F.}\ \bibnamefont {Chan}},
  \bibinfo {author} {\bibfnamefont {N.}~\bibnamefont {Wurz}},\ and\ \bibinfo
  {author} {\bibfnamefont {M.}~\bibnamefont {K{\"o}hl}},\ }\href
  {https://doi.org/10.1103/PhysRevLett.124.010403} {\bibfield  {journal}
  {\bibinfo  {journal} {Phys. Rev. Lett.}\ }\textbf {\bibinfo {volume} {124}},\
  \bibinfo {pages} {010403} (\bibinfo {year} {2020})}\BibitemShut {NoStop}%
\bibitem [{\citenamefont {Chan}\ \emph {et~al.}(2020)\citenamefont {Chan},
  \citenamefont {Gall}, \citenamefont {Wurz},\ and\ \citenamefont
  {K{\"o}hl}}]{Chan2020}%
  \BibitemOpen
  \bibfield  {author} {\bibinfo {author} {\bibfnamefont {C.~F.}\ \bibnamefont
  {Chan}}, \bibinfo {author} {\bibfnamefont {M.}~\bibnamefont {Gall}}, \bibinfo
  {author} {\bibfnamefont {N.}~\bibnamefont {Wurz}},\ and\ \bibinfo {author}
  {\bibfnamefont {M.}~\bibnamefont {K{\"o}hl}},\ }\href
  {https://doi.org/10.1103/PhysRevResearch.2.023210} {\bibfield  {journal}
  {\bibinfo  {journal} {Phys. Rev. Research}\ }\textbf {\bibinfo {volume}
  {2}},\ \bibinfo {pages} {023210} (\bibinfo {year} {2020})}\BibitemShut
  {NoStop}%
\bibitem [{\citenamefont {Jaksch}\ and\ \citenamefont
  {Zoller}(2003)}]{Jaksch2003}%
  \BibitemOpen
  \bibfield  {author} {\bibinfo {author} {\bibfnamefont {D.}~\bibnamefont
  {Jaksch}}\ and\ \bibinfo {author} {\bibfnamefont {P.}~\bibnamefont
  {Zoller}},\ }\href {https://doi.org/10.1088/1367-2630/5/1/356} {\bibfield
  {journal} {\bibinfo  {journal} {New Journal of Physics}\ }\textbf {\bibinfo
  {volume} {5}},\ \bibinfo {pages} {56} (\bibinfo {year} {2003})}\BibitemShut
  {NoStop}%
\bibitem [{\citenamefont {Aidelsburger}\ \emph {et~al.}(2011)\citenamefont
  {Aidelsburger}, \citenamefont {Atala}, \citenamefont {Nascimbene},
  \citenamefont {Trotzky}, \citenamefont {Chen},\ and\ \citenamefont
  {Bloch}}]{Aidelsburger2011}%
  \BibitemOpen
  \bibfield  {author} {\bibinfo {author} {\bibfnamefont {M.}~\bibnamefont
  {Aidelsburger}}, \bibinfo {author} {\bibfnamefont {M.}~\bibnamefont {Atala}},
  \bibinfo {author} {\bibfnamefont {S.}~\bibnamefont {Nascimbene}}, \bibinfo
  {author} {\bibfnamefont {S.}~\bibnamefont {Trotzky}}, \bibinfo {author}
  {\bibfnamefont {Y.-.~A.}\ \bibnamefont {Chen}},\ and\ \bibinfo {author}
  {\bibfnamefont {I.}~\bibnamefont {Bloch}},\ }\href
  {https://doi.org/10.1103/PhysRevLett.107.255301} {\bibfield  {journal}
  {\bibinfo  {journal} {Physical Review Letters}\ }\textbf {\bibinfo {volume}
  {107}},\ \bibinfo {pages} {255301} (\bibinfo {year} {2011})}\BibitemShut
  {NoStop}%
\end{thebibliography}

\clearpage
\onecolumngrid
\appendix

\section{Complex-time Krylov space augmented spectral functions from MPSs}
\label{supp:CTKS}
\begin{figure}
    \centering
    \includegraphics{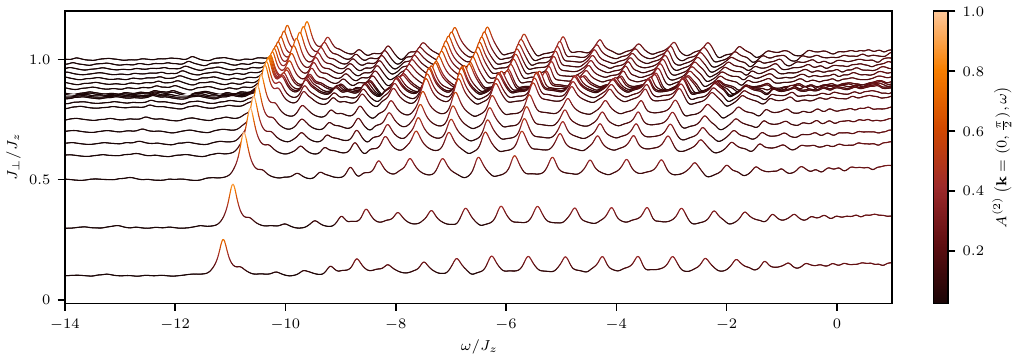}
    \caption{
        Two-hole spectral functions $A^{(2)}(\mathbf{k},\omega)$ evaluated at $\mathbf{k}=(0,\frac{\pi}{2})$ and d-wave symmetry $m_4=2$ as a function of $\omega$ and $J_\perp$.
        Using a complex-time Krylov space augmentation, we increased the frequency resolution by an order of magnitude compared to standard methods employing a real-time evolution with subsequent Fourier transformation into frequency space~\cite{Paeckel2024}.
    }
    \label{fig:spectral-function:jp-scan:kx-0}
\end{figure}
The computation of the two-hole spectral function of~\cref{eq:hamiltonian} using time-dependent \gls{MPS} is a challenging task due to the two-dimensional nature of the problem.
Typically, this is done by simulating the dynamics of time-dependent correlation functions followed by a subsequent Fourier transformation~\cite{PAECKEL2019}.
For the two-hole spectral function, this amounts to computing
\begin{equation}
    A^{(m_4)}(\mathbf k, \omega)
    =
    \operatorname{Re}\left[ \lim_{T\rightarrow\infty} \int_{0}^T \mathrm dt\, \bra{E_0} \left(\hat \Delta^{(s)}(\mathbf k;m_4)\right)^\dagger \mathrm e^{-\mathrm i(\hat H - E_0 - \omega - \mathrm i\eta)t} \hat \Delta^{(s)}(\mathbf k;m_4) \ket{E_0}\right] \;,
\end{equation}
where $\hat \Delta^{(s)}(\mathbf k;m_4)$ is the Fourier transform into momentum space of the two-hole excitation~\cref{eq:pair-creation-operator}, $\ket{E_0}$ denotes the ground state with energy $E_0$, and $\eta>0$ is a finite broadening.
In practice, the limit $T\rightarrow\infty$ is not feasible, and for the case of~\cref{eq:hamiltonian} evolution times are mostly limited to $\sim 1.66$ units of the hopping time $1/t$ (i.e., $T=5/J_z)$, especially in the Heisenberg limit.
As a consequence, the frequency resolution of the two-hole spectral function is limited by the Nyquist-Shannon theorem to $\Delta\omega = \frac{2\pi}{T} \sim 1J_z$, which is about an order of magnitude above the energy resolution required to reveal the two-channel physics.
In order to increase the frequency resolution, we exploit a complex-time Krylov-space augmentation of the spectral function introduced in~\cite{Paeckel2024}.
Here, the key idea is to rewrite the time integral into frequency space as a sum over finite-time domains of length $T$ and make use of the group property of the time-evolution operator
\begin{align}
    A^{(m_4)}(\mathbf k, \omega)
    &=
    \operatorname{Re}
    \left[
        \left(\int_0^T + \int_T^{2T} + \cdots \right) 
        \mathrm dt\, \bra{E_0}\left(\hat \Delta^{(s)}(\mathbf k;m_4)\right)^\dagger \underbrace{\mathrm e^{-\mathrm i(\hat H - E_0 - \omega - \mathrm i\eta)t}}_{:=\hat U(t,\omega+\mathrm i\eta)} \hat \Delta^{(s)}(\mathbf k;m_4) \ket{E_0}
    \right] \notag \\
    &=
    \operatorname{Re}
    \left[
        \int_0^T \mathrm dt\, \bra{E_0} \left(\hat \Delta^{(s)}(\mathbf k;m_4)\right)^\dagger \hat U(t,\omega+\mathrm i\eta) \sum_{p=0} \left(\hat U(T,\omega+\mathrm i\eta)\right)^p \hat \Delta^{(s)}(\mathbf k;m_4) \ket{E_0}
    \right] \notag \\
    &=
    \operatorname{Re}
    \left[
        \int_0^T \mathrm dt\, \bra{E_0} \left(\hat\Delta^{(s)}(\mathbf k;m_4)\right)^\dagger \hat U(t,\omega+\mathrm i\eta) \hat S(T,\omega+\mathrm i\eta) \hat \Delta^{(s)}(\mathbf k;m_4) \ket{E_0}
    \right] \;,
\end{align}
where $\hat S(T, \omega+\mathrm i\eta) = \left[1- \hat U(T,\omega+\mathrm i\eta)\right]^{-1}$, which boosts the dynamics from the interval $[0,T)$ to any interval $[pT, (p+1)T)$.
The boost operator is approximated in a Krylov-space $\mathcal K$ generated from a complex time evolution $\hat U_n = \mathrm e^{-\mathrm i \hat H (n\cdot\delta z)}$ of the initial state $\ket{\Delta} = \hat \Delta^{(s)}(\mathbf k;m_4)\ket{E_0}$:
\begin{equation}
    \mathcal K
    =
    \operatorname{span}
    \left\{
        \ket{\Delta_0}, \ket{\Delta_1}, \cdots, \ket{\Delta_{K-1}}
    \right\} \; , \quad \text{with} \quad \ket{\Delta_n} = \frac{\hat U_n \ket{\Delta}}{\lVert \hat U_n \ket{\Delta}\rVert} \; .
\end{equation}
In our simulations we used $\delta z = \delta t(1-\mathrm i\tan(\alpha))$, $\delta t = 0.1$ and $\alpha=0.03\pi$.
These values are chosen such that for the achieved Krylov-space dimension the noise threshold defined in~\cite{Paeckel2024} satisfies $R \lesssim 0.1$, which is sufficient to reliably discriminate the different two-hole states.
A set of $D\leq K$ orthonormalized basis states is obtained by computing the Gram matrix $G_{nm} = \langle \Delta_n\vert\Delta_m\rangle$ from all complex time evolved states, which permits to expand the Hamiltonian as well as the initial guess state in this new basis.
In Fig.~\ref{fig:spectral-function:jp-scan:kx-0}, we show the two-hole spectral function with d-wave symmetry ($m_4=2$) at $\mathbf{k}=(0,\pi/2)$, obtained from real-time evolutions until $T=3/J_z$ and a complex-time Krylov space with dimension $D=90$ constructed from evolving until $T=9/J_z$.
We used a broadening $\eta/J_z=0.1$, i.e., the spectral resolution $\Delta\omega \equiv \eta$ is increased by a factor of $\sim 10$ compared to a real-time evolution only, which yields $\Delta\omega/J_z = 2\pi/TJ_z \approx 2$ due to the Nyquist-Shannon sampling theorem.

\section{Effective two-channel model}
\label{app:TwoChanMdl}
The derivation of the effective model \eqref{eq:eff-model} can be split into three parts: the solution of the bare (sc) problem, the solution of the bare (cc) branch and the computation of the interchannel coupling $V_{\mathbf{k} \mathbf{k}'}^{nn'}$.

First, to describe a bare magnetic polaron (sc), we follow the method put forth in Ref.~\cite{Bermes2024} and consider a single mobile hole in an \gls{AFM}. As the hole moves through the magnetic background, it leaves behind a \emph{geometric string} of displaced spins, frustrating the magnetic order. We then build a truncated Krylov space $\{\hat{\mathcal{H}}_t^n \,\hat{c}_{i,\sigma} |\textrm{N\'eel}\rangle\}_{n\leq l_\textrm{max}}$ applying the hopping Hamiltonian to a hole in an unperturbed N\'eel background. Finally, we use numeric diagonalization to obtain the dispersion relations $\epsilon^{(\textrm{sc})}_{\mathbf{k},n}$ and corresponding eigenstates. Hence, the (sc) dispersion $\epsilon^{(\textrm{sc})}_{\mathbf{k},n}$ used in our effective model \eqref{eq:eff-model} are calculated from first principles.

Second, the tightly bound hole pairs (cc) are modeled as two holes bound by the geometric string of displaced spins. Their description is done very similarly to the single hole case, the main difference being the initial state of the truncated Hilbert space
$\{\hat{\mathcal{H}}_t^n \,\hat{c}_{i,\sigma}\hat{c}_{i+1,\overline{\sigma}} |\textrm{N\'eel}\rangle\}_{n\leq l_\textrm{max}}$,
where we consider two holes on neighboring sites in an \gls{AFM} background.
Details of this procedure are described in Ref.~\cite{Homeier2024}.
However, this method is approximate and cannot capture all effects of spin fluctuations. While the (sc) dispersion is captured very well up to a momentum-independent energy offset (as shown in Ref.~\cite{Bermes2024}), the bandwidth of the (cc) appears to be too small compared to \gls{MPS} simulations. We thus choose to use the same fit for the (cc) dispersion as in Ref.~\cite{Bohrdt2023} and reproduced in Eq.~\eqref{eq:disp-cc}.

Lastly, the coupling between the two effective channels can be understood using our geometric string framework. Here, the (cc) are two holes connected by a geometric string and spin-exchange terms $\propto J_{\perp}$ as well as next-nearest neighbor tunneling terms $\propto t'$ can break up this string, splitting the (cc) into two (sc)'s. In this article we only take the spin-exchange processes into account but both are described in more detail in Ref.~\cite{Homeier2024}. Formally, this leads to the following interchannel coupling:
\begin{equation}
    V_{\mathbf{k},\mathbf{k'}}^{nn'} = \langle 0| \,\hat{b}_{\mathbf{k}+\mathbf{k'}} \,\hat{\mathcal{H}}_{J_\perp} \hat{f}_{\mathbf{k},\uparrow,n}^{\dagger}\hat{f}_{\mathbf{k'},\downarrow,n'}^{\dagger}|0\rangle\,.
\end{equation}
Here, as in the main text, $\hat{f}_{\mathbf{k},\sigma,n}^{\dagger}$ and $\hat{b}_{\mathbf{k}}^{\dagger}$ create a (sc) and (cc) respectively, with momentum $\mathbf{k}$ and spin $\sigma = \,\uparrow, \downarrow$. Again, these coupling matrix elements are computed from first principles and are not subject to any fitting.
The evaluation of this expression is straightforward since we describe the (sc) and the (cc) in the same framework and obtain the eigenstates through numeric diagonalization.

\section{Two-hole spectrum and (cc) self-energy}
\label{app:cc-self-energy}
Starting from the two-channel model \eqref{eq:eff-model},
we define the time-ordered Green's functions at zero temperature for the (sc)
\begin{equation}
    G_{\sigma,n}(k, t) = -i\langle T \hat{f}_{k,\sigma,n}(t)\hat{f}_{k,\sigma,n}^{\dagger}(0)\rangle\,,
\end{equation}
as well as for the (cc)
\begin{equation}
    D(k, t) = -i\langle T \hat{b}_{k}(t)\hat{b}_{k}^{\dagger}(0)\rangle\,.
\end{equation}
Without interactions, the time-ordered propagators are given by
\begin{align}
    G_{n}^{(0)}(k, \omega) &= \dfrac{1}{\omega - (\epsilon^{\textrm{sc}}_{k,n}-\mu^{\textrm{sc}}) + i\textrm{sign}(\epsilon_{k,n}^{\textrm{sc}}-\mu^{\textrm{sc}})\eta}\\
    D^{(0)}(k, \nu) &= \dfrac{1}{\nu - (\epsilon^{\textrm{cc}}_{k}-\mu^{\textrm{cc}}) + i\eta}\,,
\end{align}
where $\eta \rightarrow 0^{+}$ and where we dropped the spin index.

We want to compute the (cc) spectral function $A(k,\nu)$ which is given by
\begin{equation}
    A(k, \nu) = -\dfrac{1}{\pi}\textrm{Im}(D(k,\nu + i0^{+}))\,.
\end{equation}
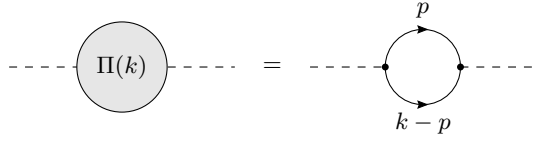
\begin{figure}[t]
    \centering
    \begin{tikzpicture}
      \coordinate (a) at (0,0);
      \coordinate (b) at (3,0);
    
      \draw[dashed] (a) -- (b);
      % Draw self-energy circle in between
      \node[draw, circle, minimum size=1.2cm, align=center, fill=gray!20] at ($(a)!0.5!(b)$) {$\Pi(k)$};
      \node at (3.5,0) {$=$};
      \draw[dashed] (4,0) -- (5,0);
      \node[circle, scale=0.3, align=center,fill=black] at (5,0) {};
      \draw (5,0) arc (180:0:0.5);
    
      \draw[-{Stealth[length=1.5mm,inset=0.2mm]}] (5.555,0.5) -- (5.575,0.5);
      \node at (5.5, 0.75) {$p$};
      \draw[-{Stealth[length=1.5mm,inset=0.2mm]}] (5.555,-0.5) -- (5.575,-0.5);
      \node at (5.5, -0.75) {$k-p$};
    
      \draw (5,0) arc (-180:0:0.5);
      \node[circle, scale=0.3, align=center,fill=black] at (6,0) {};
      \draw[dashed] (6,0) -- (7,0);
      
    \end{tikzpicture}
    \caption{The (cc) self-energy. The full lines represent the (sc) propagator and the dashed lines represent the (cc) propagator.}
    \label{fig:self-energies}
\end{figure}
The full (interacting) propagators are given by
\begin{align}
    G_{n}(k, \omega)^{-1} &= G_{n}^{(0)}(k, \omega)^{-1} - \Sigma_{n}(k, \omega)\\
    D(k, \nu)^{-1} &= D^{(0)}(k, \nu)^{-1} - \Pi(k, \nu)\,,
\end{align}
and the self-energies read
\begin{align}
    \label{eq:sc-self-energy}
    \Sigma_{n}(k, \omega) &= -i \sum_{n'}\int \dfrac{d\omega'}{2\pi}\int \dfrac{d^2p}{(2\pi)^2} |V_{k, p}^{nn'}|^2\,D(k+p, \omega + \omega')G_{n'}(p, \omega')\\
    \label{eq:cc-self-energy}
    \Pi(k, \nu) &= i \sum_{nn'}\int \dfrac{d\omega}{2\pi}\int \dfrac{d^2p}{(2\pi)^2} |V_{p, k-p}^{nn'}|^2 \, G_{n'}(k-p, \nu - \omega)G_{n}(p, \omega)\,.
\end{align}
At zero doping $\delta = 0$, it is easy to see, that the (sc) self-energy $\Sigma$ vanishes and that the magnetic polaron does not get dressed $G_{n}(k, \omega) = G_{n}^{(0)}(k, \omega)$. Thus Eq. \eqref{eq:cc-self-energy} simplifies and we can perform the frequency integration analytically.
\begin{equation}
    \label{eq:cc-self-energy-zero-doping}
    \Pi(k, \nu) = \sum_{nn'}\int \dfrac{d^2p}{(2\pi)^2} \dfrac{|V_{p, k-p}^{nn'}|^2}{\nu - \epsilon^{\textrm{sc}}_{p,n} - \epsilon^{\textrm{sc}}_{k-p,n'} + i\eta}\,.
\end{equation}
Here we used that for $\delta = 0$, $\epsilon_{k,n} - \mu \leq 0$ for all momenta $k$.

\section{Two-hole spectra in extended 2D systems}
\label{app:2d-sys}
\begin{figure}[t]
    \centering
    \includegraphics{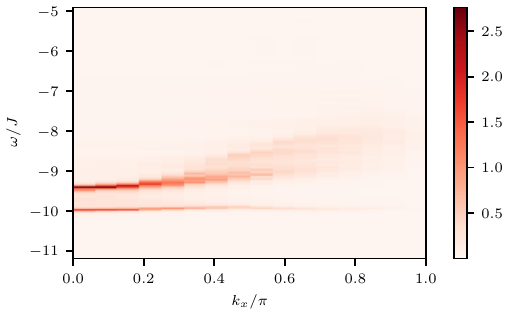}
    \caption{The d-wave ($m_4=2$) pair spectrum $A^{(2)}(\mathbf{k},\omega)$ in a two-dimensional $32\times32$ system for fixed $k_y=\pi/2$.
    For the two-channel model we used the fit parameter $\Delta E = -0.13 J$.}
    \label{fig:cc-spectrum-2D}
\end{figure}

Fig.~\ref{fig:cc-spectrum-2D} shows the same $d$-wave pair spectral function $A^{(2)}(\mathbf{k},\omega)$, computed from the effective two-channel model, as in Fig.~\ref{fig:comp_spectra_sebastian}b) in the main text; the only difference being that here we consider a two-dimensional $32\times32$ system instead of the $4$-leg cylinder considered above. As expected, the two-dimensional system features less finite-size quantization effects, which leads in this case to only two different branches in the spectrum compared to several distinguishable ones in Fig.~\ref{fig:comp_spectra_sebastian}. Note that in this case we cannot make any predictions on the fitting parameter $\Delta E$, since we don't have any two-dimensional, numerically exact data to compare to.

\section{Raman induced hopping}\label{app:raman-assisted-tunneling}
Following the proposal in Ref.~\cite{Jaksch2003}, we want to introduce laser driven Raman transitions between the states $|\uparrow\downarrow,0\rangle$ and $|\downarrow,\downarrow\rangle$. Mapping to the repulsive Hubbard model, this transitions corresponds to the creation of a hole pair from the N\'eel ordered background.
The two lasers driving this transition consist of two running plane waves yielding position-dependent Rabi frequencies
\begin{equation}
    \Omega(\mathbf{r}) = \Omega e^{i \mathbf{q}\mathbf{r}}\,,
\end{equation}
where $\mathbf{q} = \mathbf{k}_1 - \mathbf{k}_2$ denotes the difference between the wave vectors of the two plane waves.
More precisely, the two running lasers have frequencies $\omega_{1,2}$ and corresponding Rabi frequencies $\Omega_{1,2}$ for driving the transition $|\uparrow\downarrow,0\rangle \leftrightarrow|r\rangle$ and $|r\rangle \leftrightarrow|\downarrow,\downarrow\rangle$ to some excited Raman state $|r\rangle$. The detuning for the transition to the excited Raman state $\Delta_r = E_{r} - E_{\uparrow\downarrow,0} - \omega_1$ should be significantly larger than the detuning of the 2-photon process $\delta = \omega_1 - \omega_2 - E_{\downarrow,\downarrow} + E_{\uparrow\downarrow,0}$.
Upon adiabatically eliminating the auxiliary level $|r\rangle$, the resulting Rabi frequency reads
\begin{equation}
    \Omega(\mathbf{r}) = \dfrac{\Omega_{1}\Omega_{2}}{2\Delta_r}e^{i(\mathbf{k}_1 - \mathbf{k}_2)\mathbf{r}}\,.
\end{equation}
Assuming that the lasers do not excite higher Bloch bands and neglecting non-resonant transitions, the lasers lead to the additional term in the Hamiltonian
\begin{equation}
\label{eq:raman-modulation}
    \hat{\mathcal{H}}_{\textrm{las}}(\tau;\,\mathbf{q}, \delta) = \sum_{\langle \mathbf{i},\mathbf{j} \rangle}t_{\mathbf{i}\mathbf{j}}(\mathbf{q}) \,e^{i\delta  \tau}\,\hat{c}_{\mathbf{j},\downarrow}^\dagger\hat{c}_{\mathbf{i},\uparrow} + \textrm{H.c.}\,,
\end{equation}
with time $\tau$.
The hopping matrix element $t_{\mathbf{i}\mathbf{j}}(\mathbf{q})$ is given by
\begin{equation}
    t_{\mathbf{i}\mathbf{j}}(\mathbf{q}) = \dfrac{1}{2}\Omega e^{i\mathbf{q} \cdot (\mathbf{r_i} + \mathbf{r_j})/2} \int d\mathbf{r}\, \mathbf{w}^{*}(\mathbf{r}-\boldsymbol{\delta}) e^{i\mathbf{q}\mathbf{r}} \mathbf{w}(\mathbf{r}+\boldsymbol{\delta})\,,
\end{equation}
where $\boldsymbol{\delta} = (\mathbf{r_j} - \mathbf{r_i})/2$ and $\mathbf{q} = \mathbf{k}_1 - \mathbf{k}_2$ is the difference between the wave vectors of the two laser beams and $\mathbf{w}(\mathbf{r})$ is the Wannier function of the optical lattice potential.

Within linear-response theory, treating $\hat{H}_{\textrm{las}}(\tau)$ from Eq.~\eqref{eq:raman-modulation} as a time-dependent perturbation yields direct access to the two-particle spectrum after performing the particle–hole transformation in Eq.~\eqref{eq:mapping}.
Restricting the hopping amplitude to nearest-neighbor sites, $t_{\langle\mathbf{i}\mathbf{j}\rangle}(\mathbf{q}) = t_{\mathrm{eff}}(\mathbf{q})$, the modulation~\eqref{eq:raman-modulation} probes the spectral function
\begin{equation}
    A^{\textrm{Raman}}(\mathbf{k},\omega) = - \pi^{-1} {\rm Im} \mathcal{G}^{\textrm{Raman}}(\mathbf{k},\omega)\,,
\end{equation}
which is directly related to the rotational pair spectrum $A^{(m_4)}(\mathbf{k},\omega)$ discussed in the main text.
In the definition of the retarded two-particle Green's function~\eqref{eqDefG}, we simply replace the pair creation operator
\begin{equation}
    \hat{\Delta}^{(\textrm{s})}(\mathbf{k}; m_4) = \dfrac{1}{\sqrt{N}}\sum_\mathbf{j} e^{i\mathbf{k}\mathbf{j}}\sum_{\mathbf{i}:\langle\mathbf{i},\mathbf{j}\rangle}e^{im_4\varphi_{\mathbf{i} - \mathbf{j}}}(\hat{c}_{\mathbf{i},\downarrow}\hat{c}_{\mathbf{j},\uparrow} - \hat{c}_{\mathbf{i},\uparrow}\hat{c}_{\mathbf{j},\downarrow})\,,
\end{equation}
by
\begin{equation}
    \hat{\Delta}^{\textrm{Raman}}(\mathbf{k}) = \dfrac{1}{\sqrt{N}}\sum_\mathbf{j}e^{i\mathbf{k}\mathbf{j}}\sum_{\mathbf{i}:\langle\mathbf{i},\mathbf{j}\rangle}e^{i \mathbf{k}(\mathbf{j} - \mathbf{i})/2}\hat{c}_{\mathbf{j},\downarrow}\hat{c}_{\mathbf{i},\uparrow}\,.
\end{equation}
Under the partial particle-hole transformation \eqref{eq:mapping}, this maps to 
\begin{equation}
    \hat{\tilde{\Delta}}^{\textrm{Raman}}(\mathbf{k}) = \dfrac{1}{\sqrt{N}}\sum_\mathbf{j}e^{i(\mathbf{k}+(\pi,\pi))\mathbf{j}}\sum_{\mathbf{i}:\langle\mathbf{i},\mathbf{j}\rangle}e^{i \mathbf{k}(\mathbf{j} - \mathbf{i})/2}\hat{c}^\dagger_{\mathbf{j},\downarrow}\hat{c}_{\mathbf{i},\uparrow}\,.
\end{equation}
The corresponding spectrum $A^{\textrm{Raman}}(\mathbf{k},\omega)$ can directly be measured via the rate at which the pairs $|\downarrow,\downarrow\rangle$ are created. According to Fermi's golden rule, this rate is given by
\begin{align}
    \Gamma(\mathbf{k}, \omega) &= \dfrac{2\pi}{\hbar}\sum_\alpha|\langle \alpha |\hat{\mathcal{H}}_{\textrm{las}}(\tau=0;\mathbf{k},\omega)|0\rangle|^2\delta(E_\alpha - E_0 - \omega)\\
    &= \dfrac{2\pi}{\hbar}\sum_\alpha|\langle \alpha |t_\textrm{eff}\sum_{\langle\mathbf{i},\mathbf{j}\rangle}e^{i\mathbf{k} \cdot (\mathbf{r_i} + \mathbf{r_j})/2}\,
    \hat{c}_{\mathbf{j},\downarrow}^\dagger\hat{c}_{\mathbf{i},\uparrow}|0\rangle|^2\delta(E_\alpha - E_0 - \omega)\\
    &= \dfrac{2\pi |t_\textrm{eff}|^2}{2\hbar N}\sum_\alpha|\langle \alpha |\hat{\tilde{\Delta}}^{\textrm{Raman}}(\mathbf{k}+(\pi,\pi))|0\rangle|^2\delta(E_\alpha - E_0 - \omega)\\
    &=\dfrac{\pi |t_\textrm{eff}|^2}{\hbar N}A^\textrm{Raman}(\mathbf{k} + (\pi,\pi),\, \omega)\,.
\end{align}

\section{Pair Spectroscopy from Raman-induced hopping}
\label{supp:spectroscopy}
\begin{figure}
    \centering
    \includegraphics{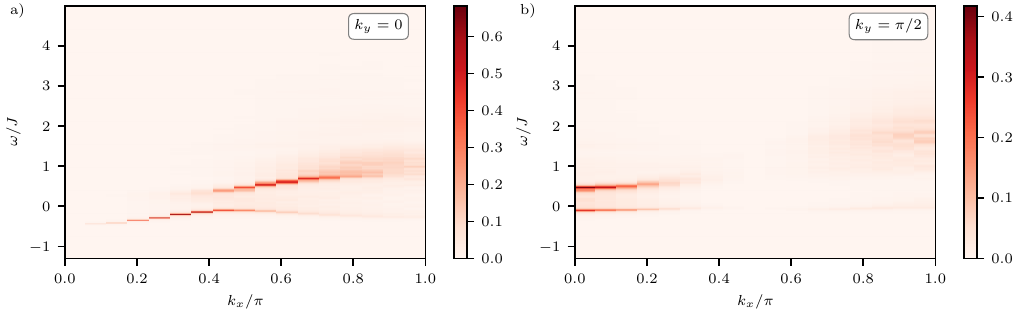}
    \caption{The spectral function of the Raman assisted tunneling modulation \eqref{eq:raman-modulation}. As in the main text, we only use the lowest cc band, set $t/J=3$ and the fit parameter $\Delta E = -0.13J$. We consider a lattice of $32\times32$ sites.}
    \label{fig:raman-spectrum}
\end{figure}
Here, we numerically compute the pair spectra measured with the Raman scheme described in Sec.~\ref{sec:experiment}. To this end, we consider a planar two-dimensional geometry in contrast to the cylinders used in the main text. We find that even though the spectral weight of the pair excitation created through Raman assisted tunneling almost vanishes for $k_x=k_y$ due to the d-wave symmetric nature of the hole-pair ground state, the avoided level crossing level crossing is still observable around the momentum points $(\pm\pi/2,0)$ and $(0, \pm\pi/2)$ as shown in Fig.~\ref{fig:raman-spectrum}.

\section{Additional momentum cuts}
\label{app:add-mom-cuts}
\begin{figure}[t]
    \centering
    \includegraphics{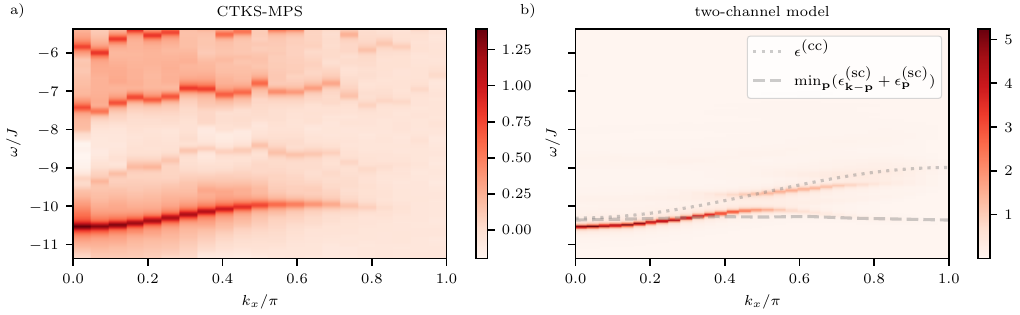}
    \caption{
    Comparison of the d-wave ($m_4 = 2$) two-hole spectrum computed numerically, a), to the effective two-channel model, b). We study the $SU(2)$ invariant $t-J$ model ($J_\perp = J_z=J$) at $k_y = 0$ as a function of $k_x$. In a) we performed high-resolution \gls{CTKS}-\gls{MPS} simulations on a $40 \times 4$ cylinder. For the two-channel model calculations in b) we used the fit parameter $\Delta E = 0.15 J$. We indicate the lower edge of the two-hole scattering continuum (dashed) and the dispersion of the tightly-bound (cc) pair (dotted).}
    \label{fig:res-comp-spectra-Annabelle}
\end{figure}

In Fig.~\ref{fig:res-comp-spectra-Annabelle}, we compare the time-dependent \gls{CTKS}-\gls{MPS} simulations to our two-channel model for an additional momentum cut with $k_y=0$.
Along the this cut, the effective model remains in qualitative agreement with the numerical simulations as it captures the curving down and the quick drop in spectral weight near $k_x=\pi/2$. However, the effective 2-channel model shows a weak second branch that is absent in the \gls{MPS} calculations.

\end{document}